\documentclass[11pt]{article}
\usepackage[utf8]{inputenc}
\usepackage[T1]{fontenc}
\usepackage{lmodern}
\usepackage{microtype}
\usepackage{inconsolata}
\usepackage{natbib}
\usepackage{soul}

\usepackage{amsmath}
\usepackage{amssymb}
\usepackage{amsfonts}

\usepackage{graphicx}
\usepackage{subcaption}
\usepackage{booktabs}
\usepackage{multirow}
\usepackage{array}
\usepackage{tabularx}
\usepackage[table]{xcolor}  % the [table] option enables \rowcolor

\usepackage[most]{tcolorbox}

\usepackage{listings}
\usepackage{algorithm}
\usepackage{algpseudocode}

\usepackage[
    colorlinks=true,
    linkcolor=blue,
    citecolor=blue,
    urlcolor=blue
]{hyperref}

\usepackage[nameinlink,noabbrev]{cleveref}

\usepackage[margin=1in]{geometry}
\usepackage{enumitem}
\usepackage{balance}

\usepackage{times}
\usepackage{latexsym}
\usepackage{microtype}
\usepackage{graphicx}
\usepackage{subcaption}
\usepackage{booktabs}
\usepackage{multirow}
\usepackage{amsmath,amssymb}
\usepackage{xspace}
\usepackage{url}
\usepackage{hyperref}
\usepackage{enumitem}
\usepackage{listings}
\usepackage{xcolor}
\usepackage[htt]{hyphenat}

\newcommand{\tech}{\textsc{ClawAudit}\xspace}

\newcommand{\openc}{OpenClaw\xspace}
\newcommand{\nanobot}{Nanobot\xspace}

\usepackage{threeparttable}
\usepackage{tablefootnote} % Must be loaded after hyperref

\usepackage{hyperref}
\hypersetup{
    colorlinks=true,
    linkcolor=blue, % Makes the footnote number clickable
    filecolor=magenta,      
    urlcolor=cyan,
}

\newtcolorbox{findingbox}[1]{
  colback=gray!6,
  colframe=black!60,
  title=\textbf{#1},
  fonttitle=\bfseries,
  boxrule=0.6pt,
  arc=2pt,
  left=6pt,
  right=6pt,
  top=5pt,
  bottom=5pt,
  before skip=8pt,
  after skip=8pt
}

\usepackage{titlesec}
\titlespacing{\paragraph}{0pt}{0.5\baselineskip}{1em}

\date{}

\title{Local LLM Agents as Vulnerable Runtimes:\\A Source-Code Audit of the Agent Runtime Layer}

\author{
Zhengsong Zhang\thanks{Equal contribution.} \quad
Zongze Li\footnotemark[1] \quad
Jiawei Guo \quad
Haipeng Cai \\
Department of Computer Science and Engineering\\
University at Buffalo, SUNY \\
\texttt{\{zhengson,zongzeli,jiaweigu,haipengc\}@buffalo.edu}
}

\begin{document}
\maketitle

\begin{abstract}
Local LLM agents such as OpenClaw and Nanobot run directly on
end-user machines and act on host resources---the shell,
filesystem, browser, stored credentials, and messaging applications---through natural-language goals. In effect, these agents have become privileged software runtimes that mediate between user intent, model outputs, and host-level actions. Existing research has characterized
this landscape through prompt injection, malicious skills, marketplace risks, or black-box behavioral evaluation of agents. But the implementation layer that performs the mediation, the prompt builder, parser, tool dispatcher, skill loader, memory writer, network client, and permission gate, has remained an unexamined safety boundary. To our knowledge, no prior work has systematically examined the agent's own source tree to audit these components for implementation-level security weaknesses.
%We present \tech, an instrument for measuring safety for such runtime layer. 
We present \tech, a static auditing framework for measuring implementation-level vulnerability exposure in the runtime layer of local LLM agents.
\tech derives a five-category
vulnerability taxonomy from the STRIDE threat model and %, for each category, 
develops custom static-analysis rules that target agent-specific patterns absent from established rule sets for vulnerability analysis. 
%implements custom static-analysis rules for the resulting runtime boundaries. 
%ClawAudit derives a five-category vulnerability taxonomy from STRIDE and implements domain-specific static-analysis rules for the resulting runtime boundaries.
We instantiate the taxonomy in two
backends, 47 Semgrep YAML rules and 30 CodeQL queries, and evaluate
both on \textsc{OpenClawBench}, a benchmark of 446 source-code-level advisories
collected from the OpenClaw repository and split temporally into 229
rule-derivation (train) and 217 held-out (test) advisories. On the held-out
test, \tech raises Semgrep recall from 21.7\% (Pro baseline) to 66.8\%, and
CodeQL recall from 13.8\% (\texttt{security-extended}) to 75.1\%. Train/test
gaps remain within 4 percentage points for all four configurations, indicating
that the rules generalize to vulnerabilities unseen during rule writing. 
% Meanwhile, a live-repository precision evaluation shows that the current rules are 
% recall-oriented and require substantial triage, motivating semantic filtering and rule 
% prioritization before deployment. 
%The recall-oriented rules trade precision for coverage, hence requiring notable post-analysis inspection effort to identify critical vulnerabilities.  
A preliminary live-code audit shows that these recall-oriented rules require substantial manual triage, motivating semantic filtering before production deployment
\end{abstract}

\section{Introduction}
\label{sec:introduction}

Local %Large Language Model (LLM) 
LLM agents, such as OpenClaw~\cite{openclaw} and Nanobot~\cite{nanobot},
operate directly on end-user machines to execute natural-language goals by
planning, reasoning, and acting. Unlike developer tools and SDKs
(e.g., LangChain~\cite{LangChain} or LlamaIndex~\cite{LlamaIndex}), these agents are end-user installable
programs with direct access to host resources, including the shell,
filesystem, browser, stored credentials, and messaging applications.

This deep system integration changes what these agents fundamentally are.
A local LLM agent is not merely a model wrapper or a tool-use policy; it is
a privileged software runtime. It accepts untrusted input, constructs
prompts, parses model output, dispatches tool calls, loads third-party
skills, writes to memory, opens network connections, and enforces
permissions, all while holding direct authority over the user's machine. 
The implementation components such as prompt builders, parsers, tool dispatchers, skill
loaders, memory writers, network clients, and permission gates are mediating between natural-language goals, model outputs, and host-level actions. 
%Thus, they are essentially becoming the actual \textbf{software runtimes} of agents, which requires strict and systematic auditing. 
Together, these components form the \textbf{software runtime} of the agent and require systematic auditing. 

However, existing works on LLM-agent safety study risks at the ecosystem or  behavioral level. One line of
work analyzes the malicious skills and unsafe agent actions in marketplace ecosystems~\cite{liu2026agent,
hasan2025model, beurer2026280+, ClawHavoc341Malicious}, while another
conducts black-box behavioral measurements of agents under emerging attacks
and defenses~\cite{greshake2023not, zhan2025adaptive, chang2025chatinject}.
Both treat the agent as a closed box and reason about its inputs and
outputs. No prior work has systematically opened the agent's own source tree to audit the runtime
components. As a 
result, this runtime layer remains an unexamined safety boundary. 

We close this gap with \tech, to our knowledge the first
systematic source-code audit of local LLM agent runtime implementations.
Guided by the STRIDE~\cite{shostack2014threat}\footnote{Spoofing identity, Tampering with data, Repudiation, Information Disclosure, Denial of Service, Elevation of Privilege~\cite{microsoft_stride}.}  threat model, we first derive a taxonomy of
implementation-level risks tailored to local agent runtimes, organized into
five domains: \textit{Prompt Handling}, \textit{Tool/Skill Execution}, \textit{Permission \&
Isolation}, \textit{Network \& Communication}, and \textit{Authentication \& Authorization}.
For each domain, we instantiate the taxonomy in two static-analysis backends:
47 custom Semgrep~\cite{semgrep} rules and 30 custom CodeQL~\cite{codeql} queries, both targeting
agent-specific patterns absent from established rule sets.
To evaluate the approach, we construct \textsc{OpenClawBench}, a benchmark of
446 publicly disclosed source-code-level advisories from the GitHub Security
Advisories (GHSA~\cite{github_advisories}) of the OpenClaw repository~\cite{openclaw}, partitioned by publication date into
229 rule-derivation (train) advisories and 217 held-out (test) advisories
collected only after the rules were finalized. On the held-out test set,
\tech raises Semgrep recall from 21.7\% (Semgrep~Pro baseline) to 66.8\%, and
CodeQL recall from 13.8\% (\texttt{security-extended} suite) to 75.1\%.
The train/test gaps remain within 4 percentage points for all four configurations. 

In summary, this paper makes the following contributions:
\begin{itemize}
    \item We identify the \emph{runtime layer} of local LLM agents as an
    under-studied safety and security boundary, distinct from prompt
    injection, malicious third-party skills, and black-box behavioral
    failures (Section~\ref{sec:background}). 
    \item We derive a STRIDE-guided taxonomy of implementation-level risks
    in local LLM agent runtimes, spanning prompt handling, tool/skill
    execution, permission and isolation, network communication, and
    authentication and authorization (Section~\ref{sec:taxonomy}).
    \item We present \tech, a domain-specific static auditing framework that
    instantiates this taxonomy in two static-analysis backends: 47 Semgrep
    YAML rules and 30 CodeQL queries (Section~\ref{sec:design}).
    \item We construct \textsc{OpenClawBench}, a benchmark of 446 disclosed
    source-code-level OpenClaw advisories with a temporal train/test split,
    and evaluate \tech against both a Semgrep~Pro baseline and a CodeQL
    \texttt{security-extended} baseline (Section~\ref{sec:benchmark}).
    % \item We characterize the boundary of rule-based runtime auditing:
    % domain-specific rules substantially improve detection for syntactically
    % structured flaws, whereas semantic authorization and
    % information-exposure flaws remain difficult.

    \item We characterize both the promise and the current deployment cost of rule-based runtime auditing: domain-specific rules substantially improve recall on disclosed advisories, but live-code precision remains low without semantic filtering (Section~\ref{sec:evaluation}).
\end{itemize}

\section{Background}
\label{sec:background}

\subsection{Local LLM Agents as Privileged Runtimes}
\label{sec:local-agent-runtime}

Local LLM agents differ from developer frameworks and SDKs in a way that is easy to overlook but central to their risk
profile. Those frameworks are libraries: a developer imports them, supplies
the surrounding application, and decides what privileges the resulting
program holds. Local agents such as OpenClaw~\cite{openclaw}, Nanobot~\cite{nanobot}, and PicoClaw~\cite{picoclaw} invert
this relationship. They ship as complete, end-user installable programs that
already contain the planning loop, the tool integrations, and the host
bindings, and they run with the full privileges of the user who launches
them.

This makes a local agent a \textit{software runtime} in the conventional system sense: a
long-lived process that ingests untrusted input, transforms it through
several internal stages, and emits privileged actions on the host. The input
is a natural-language goal; the privileged actions include executing shell
commands, reading and writing arbitrary files, driving a web browser,
sending and receiving messages, and using stored credentials such as API
keys and passwords. Between input and action lies a pipeline of
implementation components, the prompt builder, parser, tool dispatcher,
skill loader, memory writer, network client, and permission gate, each of
which inspects or rewrites data on its way to the host. We refer to this
pipeline collectively as the \ul{agent runtime}. Crucially, every stage in
it executes with the same host privileges as the agent as a whole, so the
runtime offers no internal privilege boundary that would contain a fault in
any single component.

\subsection{Existing Agent-Safety Perspectives}
\label{sec:existing-perspectives}

The risks of LLM agents have drawn growing attention, and current efforts
cluster around three concerns. The first is the \emph{model}: prompt
injection and jailbreaks, where adversarial texts hidden in instructions,
retrieved documents, or tool results steer an agent into unintended
actions~\cite{greshake2023not, zhan2025adaptive, chang2025chatinject}. The
second is \emph{third-party content}: the skills and marketplaces that local
agents draw on, where malicious entries appear at scale, over 300 malicious
skills on repositories such as ClawHub~\cite{ClawHavoc341Malicious}, and
hundreds more that exfiltrate credentials through the model
context~\cite{beurer2026280+}. The third is the \emph{agent as a whole},
treated as a black box and measured by its end-to-end behavior in
deployment, where tens of thousands of instances have been found exposed
online and open to remote code execution~\cite{OpenClawSecurityRisks}.

What these concerns share is their vantage point: each reasons about the
agent through its inputs, prompts and skills, or its outputs, observed
behavior, while taking the code that mediates between them to be correct.
The implementation that performs that mediation---the runtime (as defined in
Section~\ref{sec:local-agent-runtime})---has not itself been treated as an
object of security analysis.

\subsection{Why Runtime Auditing Matters}
\label{sec:why-runtime}

A local agent's safety is usually argued one component at a time: the model
is aligned, the skills are vetted, the sandbox confines the shell. None of
these arguments covers the code that sits between the components. The agent
runtime is what parses an instruction into a structured intent, assembles
that intent into a prompt, reads the model's reply, selects a tool, marshals
its arguments, writes the result into memory, and checks a permission before
the call goes through. A model that refuses a dangerous request does not help
if the parser strips the refusal; a benign skill does not help if the loader
runs its setup hook before the permission gate; a sandbox does not help if
the dispatcher resolves the tool path outside it. The components can each be
correct and the agent still be unsafe, because what matters is whether the
runtime preserves the guarantees the components assume.

These are ordinary software faults: an unsanitized prompt builder may be an
injection sink, a parser that evaluates a model-emitted argument may correspond to CWE-94 (\textit{Improper Control of Generation of Code ('Code Injection')}), % CWE-94, 
a skill loader that unpickles a marketplace payload may correspond to CWE-502 (\textit{Deserialization of Untrusted Data}), and a permission
gate placed after its side effect may be a check that runs too late. What makes
them different from the same bugs in an ordinary program is the authority
behind them. The runtime holds the shell, the filesystem, and the user's
stored credentials, so a missing check is not a contained defect but a direct
path from untrusted input to a privileged action on the host. 
%The marketplace skills and exposed deployments from Section~\ref{sec:existing-perspectives} are load on exactly this layer: the input arrives from outside, but the severity of what follows is decided inside the runtime, 
The marketplace skills and exposed deployments discussed in Section~\ref{sec:existing-perspectives} place pressure on exactly this layer: the input arrives from outside, but the severity of what follows is decided inside the runtime, 
by whether the loader inspects what it runs and the
dispatcher constrains what it invokes. That is the layer prior work leaves
unaudited, and the layer this paper aims to address.

\section{Agent Runtime Vulnerability Taxonomy}
\label{sec:taxonomy}

As noted earlier (Section~\ref{sec:background}), the agent runtime is a privileged
layer whose implementation has gone unaudited. To audit it, we first need to
know which weaknesses to audit for. This section answers that question by
building a taxonomy of the implementation-level vulnerabilities that can arise
in a local agent runtime: we derive the categories from a threat model rather
than naming them by intuition (Section~\ref{sec:taxonomy-stride}), then define
each one in terms of the runtime component it implicates
(Section~\ref{sec:taxonomy-categories}). The result is the set of targets the
rest of the paper aims to detect.

\subsection{Deriving the Taxonomy from STRIDE}
\label{sec:taxonomy-stride}

Auditing a class of software that has not been audited before raises a
prior question: ``\textit{what should we look for}"? Enumerating vulnerability categories
by intuition risks both omission and overlap, so we instead derive them from
STRIDE~\cite{microsoft_stride}, a threat-modeling framework~\cite{shostack2014threat} that classifies threats into six
categories 
%(Spoofing, Tampering, Repudiation, Information Disclosure, Denial of Service, and Elevation of Privilege) 
and applies them to each trust
boundary in a system's data-flow diagram. STRIDE gives us a principled way to
move from the runtime's structure to the weaknesses that structure admits.

We begin by constructing the data-flow diagram of the OpenClaw agent runtime.
Three trust boundaries dominate it. The first is where user input enters
the agent
core. The second is where the core hands work to tool and skill execution.
The third is where the agent reaches out to external resources over the
network. These are the points where the runtime takes in something it does
not control and turns it into something privileged, which is exactly where
STRIDE directs attention.

At each boundary we asked three questions: (i) \textit{what attacker-controlled data
crosses it}, (ii) \textit{what privileged operation follows once the data is across}, and
(iii) \textit{what implementation component mediates that transition}. The first question
locates the untrusted input, the second locates the consequence, and the
third names the runtime component an auditor would actually inspect. Answering
these three questions at the three boundaries is what turns an abstract threat
model into a set of concrete code-level targets, and it is the step that ties
the taxonomy to the runtime components defined in
Section~\ref{sec:local-agent-runtime}. %rather than to threats in the abstract.

Applying STRIDE this way yields four of the six threat categories. We exclude
Denial of Service (DoS) because it manifests as resource exhaustion observable only
at runtime, which a static audit of the source tree cannot reach.\footnote{We exclude generic Denial-of-Service (DoS) behavior that requires runtime workload observation. However, we retain advisories labeled CWE-400 when the disclosed fix exposes a statically visible implementation pattern, such as missing size checks or unbounded parsing. These cases are evaluated as implementation weaknesses under the closest runtime boundary rather than as a separate DoS category.} We exclude
Repudiation because it concerns the integrity of audit logging, a concern
orthogonal to the implementation weaknesses we target. The four remaining
threats (Tampering, Elevation of Privilege, Information Disclosure, and
Spoofing) map onto concrete bug categories. Tampering alone splits into two,
because attacker-controlled data tampers with the runtime at two structurally
different points: once when untrusted text enters prompt construction, and
again when untrusted data steers tool and skill execution. The components
differ, the vulnerability patterns differ, and an auditor writing rules for
one would not catch the other, so we keep them separate. This gives a final
taxonomy of five categories, summarized in Table~\ref{tab:taxonomy}.

% \begin{table}[htbp]
% \centering
% \caption{Mapping from STRIDE threats to the five runtime vulnerability categories}
% \label{tab:taxonomy}
% \begin{tabularx}{\linewidth}{l l X}
% \toprule
% STRIDE Threat & Category & Examples \\
% \midrule
% Tampering & CAT-1: Prompt Handling &
% memory contamination \\

% Tampering & CAT-2: Tool/Skill Execution &
% Path traversal, command injection \\

% Elevation of Privilege & CAT-3: Permission \& Isolation &
% Missing sandboxing \\

% Information Disclosure & CAT-4: Network \& Communication &
% SSRF, credential leakage in logs \\

% Spoofing & CAT-5: Authentication \& Authorization &
% Missing auth, privilege escalation \\
% \bottomrule
% \end{tabularx}
% \end{table}

\begin{table}[htbp]
\centering
\caption{Mapping from STRIDE threats to the five agent runtime vulnerability categories}
\label{tab:taxonomy}
\small
\setlength{\tabcolsep}{7pt}
\renewcommand{\arraystretch}{1.35}
\begin{tabularx}{\linewidth}{@{}l l X@{}}
\toprule
\rowcolor{gray!12}
\textbf{STRIDE Threat} & \textbf{Vulnerability Category} & \textbf{Representative Examples} \\
\midrule
\multirow{2}{*}{Tampering}
  & \cellcolor{blue!8}\textbf{\textcolor{blue!55!black}{CAT-1}}~Prompt Handling
  & Memory contamination \\
  & \cellcolor{teal!10}\textbf{\textcolor{teal!55!black}{CAT-2}}~Tool/Skill Execution
  & Path traversal, command injection \\
\addlinespace[3pt]
Elevation of Privilege
  & \cellcolor{orange!10}\textbf{\textcolor{orange!60!black}{CAT-3}}~Permission \& Isolation
  & Missing sandboxing \\
\addlinespace[3pt]
Information Disclosure
  & \cellcolor{red!8}\textbf{\textcolor{red!55!black}{CAT-4}}~Network \& Communication
  & SSRF, credential leakage in logs \\
\addlinespace[3pt]
Spoofing
  & \cellcolor{violet!10}\textbf{\textcolor{violet!60!black}{CAT-5}}~Authentication \& Authorization
  & Missing auth, privilege escalation \\
\bottomrule
\end{tabularx}
\end{table}

%\subsection{Taxonomy Categories}
\subsection{Resulting Taxonomy}
\label{sec:taxonomy-categories}

Each category corresponds to one of the three questions answered at a trust
boundary: an untrusted input, a privileged operation it reaches, and the
component that should have stood between them.

\textbf{CAT-1: Prompt Handling.} Untrusted text enters prompt construction or
memory. The agent assembles its prompt from user goals, retrieved documents,
tool outputs, and stored memory, and any of these can carry instructions the
developer never intended the model to follow. The mediating components are the
prompt builder and the memory writer. A weakness here lets external text reach
the model as if it were trusted instruction, whether in the current turn
through injection or in a later turn through contaminated memory.

\textbf{CAT-2: Tool/Skill Execution.} Model- or user-controlled data
influences the filesystem, the shell, or skill loading. Once the model decides
to act, the runtime turns its output into concrete operations: it resolves a
path, builds a shell command, or loads a skill module. The mediating
components are the tool dispatcher and the skill loader. A weakness here lets a
crafted argument or a malicious skill escape the intended operation, producing
the path traversal, command injection, and unsafe-load patterns familiar from
ordinary software but reached here through model output.

\textbf{CAT-3: Permission \& Isolation.} The runtime fails to isolate tools,
skills, or execution contexts from one another or from the host. Even when an
individual operation is well formed, the runtime is responsible for confining
it: sandboxing a skill, scoping a tool's filesystem access, separating one
execution context from the next. The mediating component is the permission
gate together with whatever isolation mechanism the agent provides. A weakness
here is a missing or overly broad confinement, so that a benign-looking
operation reaches further into the host than it should.

\textbf{CAT-4: Network \& Communication.} The runtime constructs requests,
logs, or external interactions unsafely. Agents fetch URLs, call APIs, and
write logs, and each of these is a place where internal data can leak outward
or where attacker-influenced input can redirect an outbound request. The
mediating component is the network client. A weakness here produces
server-side request forgery (SSRF) when a model-supplied URL is fetched without
restriction, or credential disclosure when secrets are written into logs or
sent to the wrong endpoint.

\textbf{CAT-5: Authentication \& Authorization.} The runtime fails to enforce
authorization over a user, session, or resource. Agents that expose interfaces
or manage multiple sessions must check who is asking before acting, and must
bind a resource to the session entitled to it. The mediating component is the
permission gate at the point of entry rather than at the point of action. A
weakness here is a missing or bypassable check that lets an unauthenticated or
under-privileged caller obtain access they should not have.

%\subsection{The Five Categories as Trust Boundaries}
\subsection{Formulating Trust Boundaries around the Taxonomy}
\label{sec:taxonomy-boundaries}

The five categories above, derived from STRIDE, also admit a complementary
reading as the trust boundaries crossed by data on its way through the agent
runtime. Each category $C_i$ corresponds to a boundary $B_i$ where a particular
class of untrusted data meets a particular class of privileged operation. 
Each boundary corresponds to a sanitization barrier, guarding a distinct \textbf{sink} class and answering one yes/no \textbf{question} at runtime.

% \begin{description}[leftmargin=*,nosep,labelindent=0pt,itemsep=2pt]
%     \item[$B_1$ Pre-prompt Sanitization (CAT-1).]
%           Sink: prompt builder, memory writer.
%           \emph{Question:} Is this text safe to place in trusted instructions?
%     \item[$B_2$ Operand-to-Execution (CAT-2).]
%           Sink: exec/spawn/eval, dynamic require, skill loader.
%           \emph{Question:} Is this operand safe to pass to the executor?
%     \item[$B_3$ Action-to-Resource Confinement (CAT-3).]
%           Sink: filesystem operation, sandbox operation, parser-then-allocator,
%           interface bind.
%           \emph{Question:} Does this action stay inside its sandbox / path /
%           size budget?
%     \item[$B_4$ Action-to-Outbound Egress (CAT-4).]
%           Sink: HTTP/WS request, log emission, error formatter.
%           \emph{Question:} Is this URL / log line / error message safe to emit?
%     \item[$B_5$ Caller-to-Handler Authorization (CAT-5).]
%           Sink: handler entry, route mount, skill-install gate.
%           \emph{Question:} Is this caller allowed to invoke this handler?
% \end{description}

\begin{description}[leftmargin=*,labelindent=0pt,style=nextline,
                    itemsep=8pt,parsep=1pt,topsep=4pt,labelsep=0pt]
  \item[$B_1$\enspace Pre-prompt Sanitization~%
        \textcolor{blue!55!black}{\normalfont\small(CAT-1)}]
        \textit{Sinks.}\enspace prompt builder; memory writer.\\
        \textit{Question.}\enspace Is this text safe to place in trusted instructions?
  \item[$B_2$\enspace Operand-to-Execution~%
        \textcolor{teal!55!black}{\normalfont\small(CAT-2)}]
        \textit{Sinks.}\enspace \texttt{exec}/\texttt{spawn}/\texttt{eval}; dynamic \texttt{require}; skill loader.\\
        \textit{Question.}\enspace Is this operand safe to pass to the executor?
  \item[$B_3$\enspace Action-to-Resource Confinement~%
        \textcolor{orange!60!black}{\normalfont\small(CAT-3)}]
        \textit{Sinks.}\enspace filesystem operation; sandbox operation; parser-then-allocator; interface bind.\\
        \textit{Question.}\enspace Does this action stay inside its sandbox / path / size budget?
  \item[$B_4$\enspace Action-to-Outbound Egress~%
        \textcolor{red!55!black}{\normalfont\small(CAT-4)}]
        \textit{Sinks.}\enspace HTTP/WS request; log emission; error formatter.\\
        \textit{Question.}\enspace Is this URL / log line / error message safe to emit?
  \item[$B_5$\enspace Caller-to-Handler Authorization~%
        \textcolor{violet!60!black}{\normalfont\small(CAT-5)}]
        \textit{Sinks.}\enspace handler entry; route mount; skill-install gate.\\
        \textit{Question.}\enspace Is this caller allowed to invoke this handler?
\end{description}

This boundary view supports a simple decision procedure for classifying any
runtime weakness: walk through $B_1$--$B_5$ in order, and the first sink that
matches the missing guard determines the category. We applied this procedure
when classifying each of the 47 Semgrep rules and 30 CodeQL queries; every
rule falls under exactly one boundary, which makes the taxonomy
mutually-exclusive at the level of rules. Edge cases that look ambiguous
(\emph{e.g.}, untrusted-header-into-log, which involves user input but is
fixed at the log call site, hence CAT-4 rather than CAT-1) resolve cleanly
under this rule.

\subsection{Taxonomy Scope}
\label{sec:taxonomy-scope}

This taxonomy is not intended to exhaust all possible agent failures; it
captures the implementation-level weakness classes observable in disclosed
OpenClaw advisories and relevant to static auditing. Failures that arise only
at runtime, that depend on the model's semantic judgment, or that live outside
the runtime in the model weights or the surrounding deployment fall outside
its scope, and we return to several of these as limitations in
%Section~\ref{sec:why-runtime}.
Section~\ref{sec:limitations}.

\begin{figure}[htbp]
\centering
\includegraphics[width=0.99\linewidth]{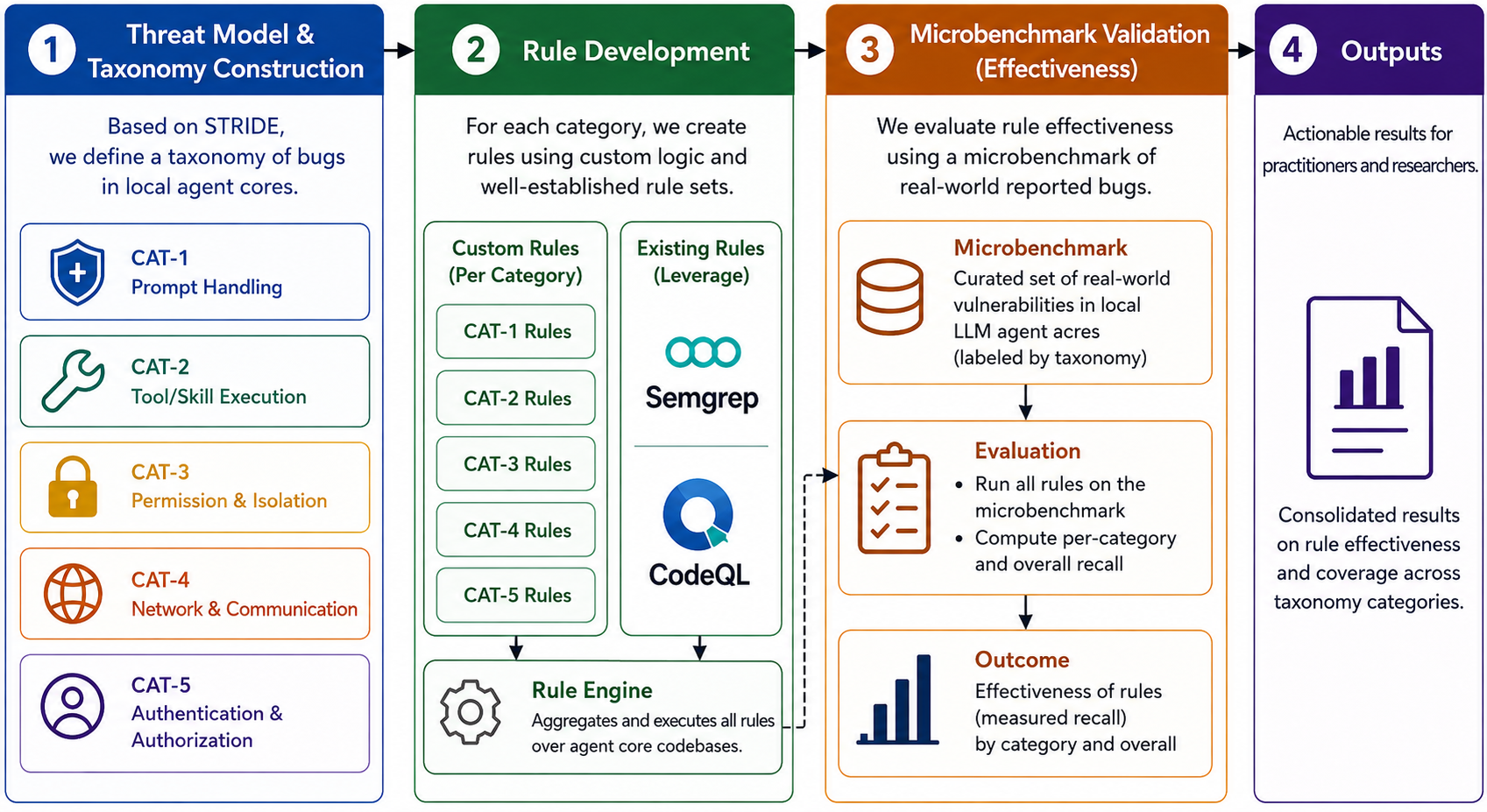}
\caption{Overview of %\tech. 
{\tech}’s taxonomy$\rightarrow$rule$\rightarrow$benchmark pipeline. 
The framework derives an agent-runtime taxonomy, encodes domain-specific static-analysis rules, and evaluates them against disclosed source-code-level advisories.}
\label{fig:overview}
\end{figure}

\section{\tech Design}
\label{sec:design}

\subsection{Overview}
\label{sec:design-overview}

As shown in Figure \ref{fig:overview}, \tech is a static auditing framework for local LLM agent runtimes. It consists of three stages:
\begin{enumerate}[leftmargin=*]
    \item Taxonomy-guided vulnerability modeling (Section~\ref{sec:taxonomy});
    \item Domain-specific rule construction in two static-analysis backends,
          Semgrep (YAML pattern rules) and CodeQL (Datalog queries over a
          relational program representation);
    \item Benchmark-based validation against disclosed advisories using a
          temporal train/test split.
\end{enumerate}
The two backends share the same taxonomy and the same target patterns but
differ in their analysis model: Semgrep matches syntactic templates over the
AST and is well suited to dense pattern enumeration, whereas CodeQL evaluates
declarative predicates over a relational program graph and supports limited
inter-procedural data-flow reasoning. Implementing the same five-category
taxonomy in both backends lets us measure how much of \tech's improvement
comes from the taxonomy itself versus the underlying analysis engine.

\subsection{Rule Development}
\label{sec:rule-development}

For each taxonomy category, we inspect vulnerable code snapshots from disclosed
advisories (Section~\ref{sec:data-collection}) and identify recurring
implementation structures. We then encode each structure twice, as a Semgrep
YAML rule and as a CodeQL query, so that every \tech rule has parallel
implementations in both backends. 

\begin{itemize}
    \item The Semgrep rule expresses the pattern as a
syntactic template over the AST with metavariable regular expressions
constraining identifier names and \texttt{pattern-not-inside} clauses excluding
common safe surroundings. 

    \item The CodeQL query expresses the same pattern as a
declarative predicate over the program's relational representation, typically
including a small auxiliary predicate that captures the absence of a
sanitizing guard in the enclosing function. 
\end{itemize}
Each rule's metadata cites at
least one ground-truth advisory used to derive it.

% \begin{table}[t]
% \centering
% \caption{Rule distribution by taxonomy category, instantiated in both
% the Semgrep and CodeQL backends.}
% \label{tab:rules}
% \begin{tabularx}{\linewidth}{X X c c}
% \toprule
% Category & Rule Families & \# Semgrep & \# CodeQL \\
% \midrule
% CAT-1: Prompt Handling
% & Prompt template, memory writer, log poisoning
% & 3 & 4 \\

% CAT-2: Tool/Skill Execution
% & Command injection, env var injection, SSH spoof
% & 5 & 3 \\

% CAT-3: Permission \& Isolation
% & Path traversal, symlink TOCTOU, sandbox, file mode
% & 14 & 12 \\

% CAT-4: Network \& Communication
% & SSRF, credential leak in logs/errors, hardcoded secrets
% & 13 & 7 \\

% CAT-5: Auth \& Authorization
% & Auth-default-allow, allowlist gap, handler bypass
% & 9 & 4 \\

% Web XSS (uncategorized\footnote{The Web XSS rules are retained as auxiliary rules because several advisories expose prompt-rendering or UI-rendering surfaces, but they are not part of the five-category runtime taxonomy.})
% & JSON-in-inline-script HTML escape
% & 3 & -- \\

% \midrule
% Total & -- & \textbf{47} & \textbf{30} \\
% \bottomrule
% \end{tabularx}
% \end{table}

\begin{table}[t]
\centering
\caption{Rule distribution by taxonomy category, instantiated in both the Semgrep and CodeQL backends. The five runtime categories contain 44 Semgrep rules and 30 CodeQL queries; three additional Web XSS rules are retained as auxiliary Semgrep-only coverage}
\label{tab:rules}
\small
\setlength{\tabcolsep}{6pt}
\renewcommand{\arraystretch}{1.45}
\begin{tabularx}{\linewidth}{@{}l X r r@{}}
\toprule
\rowcolor{gray!12}
\textbf{Category} & \textbf{Rule Families} & \textbf{\# Semgrep} & \textbf{\# CodeQL} \\
\midrule
\cellcolor{blue!8}\textbf{\textcolor{blue!55!black}{CAT-1}}~Prompt Handling
  & Prompt template; memory writer; log poisoning
  & 3 & 4 \\
\addlinespace[2pt]
\cellcolor{teal!10}\textbf{\textcolor{teal!55!black}{CAT-2}}~Tool/Skill Execution
  & Command injection; env-var injection; SSH spoof
  & 5 & 3 \\
\addlinespace[2pt]
\cellcolor{orange!10}\textbf{\textcolor{orange!60!black}{CAT-3}}~Permission \& Isolation
  & Path traversal; symlink TOCTOU; sandbox; file mode
  & 14 & 12 \\
\addlinespace[2pt]
\cellcolor{red!8}\textbf{\textcolor{red!55!black}{CAT-4}}~Network \& Communication
  & SSRF; credential leak in logs/errors; hardcoded secrets
  & 13 & 7 \\
\addlinespace[2pt]
\cellcolor{violet!10}\textbf{\textcolor{violet!60!black}{CAT-5}}~Auth \& Authorization
  & Auth-default-allow; allowlist gap; handler bypass
  & 9 & 4 \\
\addlinespace[2pt]
\cellcolor{gray!10}\textit{Web XSS}\textsuperscript{$\dagger$}
  & JSON-in-inline-script HTML escape
  & 3 & -- \\
\midrule
\rowcolor{gray!15}
\textbf{Total} & & \textbf{47} & \textbf{30} \\
\bottomrule
\multicolumn{4}{@{}p{\linewidth}@{}}{\footnotesize\textsuperscript{$\dagger$}\,The Web XSS rules are retained as auxiliary rules because several advisories expose prompt-rendering or UI-rendering surfaces, but they are not part of the five-category runtime taxonomy.}
\end{tabularx}
\end{table}

Table~\ref{tab:rules} summarizes the resulting rule set. The main taxonomy accounts for 44 Semgrep rules and all 30 CodeQL queries across the five runtime categories. CAT-3 and CAT-4 receive the largest number of rules because filesystem confinement, sandboxing, SSRF, and credential-leak patterns appear in multiple syntactic forms across the disclosed advisories. In addition to the five taxonomy categories, we retain three auxiliary Web XSS rules in the Semgrep backend. These rules cover advisories involving JSON embedded in inline scripts or UI-rendered agent state; they are useful for benchmark coverage but are not counted as part of the five-category runtime taxonomy.

% \subsection{Detection Semantics}
% \label{sec:detection-semantics}

% %A rule firing is considered a positive detection when it occurs within the patched region or the enclosing function associated with a disclosed advisory. We use advisory-level recall as the primary evaluation metric. 

% A rule firing is considered a (positive) advisory-level detection when it occurs in one of the files modified by the advisory’s fix commit. We use file-level attribution because fixes sometimes restructure the vulnerable function, making patched-region or line-level matching brittle. We consider this attribution choice a limitation of our current {\tech} design.  
% %File-level attribution may over-count detections when a modified file contains multiple unrelated risky patterns.
% File-level attribution may over-count detections when a modified file contains unrelated risky patterns, and may under-count detections when the true root cause lies outside the files modified by the fix commit. We use this attribution as a recall-oriented approximation because line-level matching is brittle when patches restructure vulnerable functions. 

\subsection{Advisory-Level Detection and Attribution}
\label{sec:detection-semantics}

\tech produces raw static-analysis findings rather than advisory-level labels. A raw finding is a tuple consisting of a rule identifier, file path, line number, and matched code snippet. To evaluate whether such findings recover disclosed vulnerabilities, we define an attribution rule that maps raw findings to advisories in \textsc{OpenClawBench}.

For each advisory, we identify the files modified by its fix commit. A configuration is considered to detect the advisory if it emits at least one finding in one of these modified files when run on the vulnerable revision of the agent runtime. We use this file-level attribution for two reasons. First, many fixes restructure the vulnerable function, move validation logic into helpers, or modify call sites rather than the original sink, making line-level or hunk-level matching brittle. Second, advisory patches often touch the nearest repair location rather than every location causally involved in the vulnerability, so requiring a finding to fall exactly inside the edited hunk would under-count detections that correctly flag the vulnerable operation nearby.

This attribution rule is intentionally recall-oriented. It asks whether a rule set points the auditor to a file implicated by the disclosed fix, not whether the exact vulnerable expression has been localized with proof-level precision. The choice has two limitations. It may over-count detections when a modified file contains unrelated risky patterns that are not the vulnerability fixed by the advisory. It may also under-count detections when the true root cause lies outside the files modified by the fix commit, for example when the patch repairs a downstream guard while the analyzer reports an upstream source. We therefore treat file-level attribution as an approximation suitable for comparing recall across rule sets, and separately evaluate live-code finding volume and manual precision in Section~\ref{sec:rq4} to characterize triage cost.

\vspace{-8pt}
\section{\textsc{OpenClawBench}}
\label{sec:benchmark}
\vspace{-4pt}
\subsection{Data Collection}
\label{sec:data-collection}

We construct \textsc{OpenClawBench} from publicly disclosed GitHub Security
Advisories (GHSAs) associated with the \openc repository. Starting from the
GitHub Security Advisories REST API, we fetched all 602 published advisories
covering the period of 2026-01-31 through 2026-05-15. 

For each advisory, we
recovered the fix commit by joining against the {\tt OSV.dev} database (which
exposes fix-commit references that the GitHub advisory API does not). We then
used the parent of the fix commit as the vulnerable revision and downloaded
each modified file at that revision. After excluding 130 advisories whose
fix commits could not be recovered, and 26 whose snapshots contained no
scannable JavaScript or TypeScript, we retained 446 advisories with vulnerable
source-code snapshots. For each resulting advisory we record the advisory identifier,
the vulnerable source-code snapshot, the modified file set, the
CWE label, the severity level, and the publication date.

\subsection{Temporal Train/Test Split}
\label{sec:split}

To measure generalization to vulnerabilities unseen during rule development,
we partition the 446 advisories temporally rather than randomly. We fix
2026-04-01 as the cutoff: 229 advisories published before the cutoff form the
\emph{rule-derivation set} (train), and 217 advisories published on or after
the cutoff form the \emph{held-out test set}. Rule development used only the
229 train advisories, and the 217 test advisories were collected only after
all rules had been committed. The held-out test set is evaluated exactly once,
without iteration, mirroring real-world deployment where rules are written from
past disclosures and applied to future, never-seen vulnerabilities. 
Also, no advisory text, vulnerable snapshot, patch diff, or file path from the test partition was inspected during rule writing. 
The CWE distribution of test advisories was not known during rule development either. 

\subsection{Benchmark Statistics}
\label{sec:benchmark-structure}

\paragraph{Severity, CWE, and category labels.}
Every advisory in \textsc{OpenClawBench} carries three labels we treat as
ground truth: a severity bucket, a CWE class, and a taxonomy category. The
severity bucket and CWE class come directly from the advisory's GHSA
metadata as assigned by the OpenClaw maintainers and the GitHub Security
Advisories reviewers; we do not re-classify them. The four severity buckets
correspond to CVSS v3 score ranges, namely \emph{critical}
(CVSS $\ge 9.0$), \emph{high} (CVSS $7.0$--$8.9$), \emph{medium}
(CVSS $4.0$--$6.9$), and \emph{low} (CVSS $0.1$--$3.9$); their CVSS
provenance makes them comparable across advisories without requiring our
own judgment. The taxonomy category is the one extra label we attach,
assigned by walking the boundary decision procedure of
Section~\ref{sec:taxonomy-boundaries}.

\paragraph{Distributions.}
\textsc{OpenClawBench} contains 446 advisories with the following
distributions (train + test combined):
12 critical, 134 high, 250 medium, and 50 low severity.
The top CWE classes are
CWE-863 authorization bypass (77),
CWE-22 path traversal (31),
CWE-78 command injection (30),
CWE-918 SSRF (20),
CWE-285 missing authorization (18),
CWE-400 resource exhaustion (17),
CWE-184 incomplete denylist (16),
CWE-59 symlink following (15),
CWE-639 user-controlled resource identifier (10),
and CWE-200 information exposure (10).
The benchmark spans 84 distinct CWE classes in total. The full long-tail
distribution and the train/test breakdown are given in
Appendices~\ref{app:benchmark} and \ref{app:cwe}.

\subsection{Evaluation Protocol}
\label{sec:evaluation-protocol}

\paragraph{Tool configurations.}
We compare four configurations across two backends:
\begin{itemize}[leftmargin=*]
    \item \textbf{Semgrep Baseline}: Semgrep Pro with \texttt{--config auto},
          covering ${\approx}2{,}815$ built-in and community rules.
    \item \textbf{Semgrep + \tech}: the Semgrep baseline augmented with
          \tech's 47 domain-specific (i.e. agent-runtime-specific) YAML rules.
    \item \textbf{CodeQL Baseline}: the \texttt{javascript-security-extended}
          query suite, covering 104 built-in queries.
    \item \textbf{CodeQL + \tech}: the CodeQL baseline augmented with
          \tech's 30 domain-specific (i.e. agent-runtime-specific) queries.
\end{itemize}

\paragraph{Metrics and measures.}
A finding $f$ is a tuple (\texttt{rule-id}, \texttt{file-path},
\texttt{line-number}, \texttt{snippet}) emitted by either backend on a given
input revision. An advisory $a$ is considered \emph{detected} by a
configuration $C$ if $C$ produces at least one finding $f$ whose
\texttt{file-path} matches one of the files modified by the fix commit of
$a$. The matching is at the file level rather than the line level: a
finding anywhere in a modified file counts, but a finding in an unrelated
file does not. We use this attribution because the patches in
\textsc{OpenClawBench} sometimes restructure a function during the fix, so
line-level matching against the patched region would systematically
under-count detections that the rule emits a few lines before or after the
edited block. Advisory-level recall is then the fraction of advisories
detected by $C$, computed separately on the train and test partitions.

\paragraph{Tool versions and hardware.}
We run Semgrep Pro version 1.142.0 (released 2026-02) and the CodeQL CLI
version 2.23.5 (released 2025-12) with the bundled \texttt{javascript-all}
extractor. All evaluations are performed on a single Linux x86\_64 host
with 64 logical cores, 256~GB of system memory, and local NVMe storage. The
host is otherwise idle during the scans.

\paragraph{Wall-clock budgets.}
For per-advisory scans on the 446 \textsc{OpenClawBench} snapshots, each
Semgrep run uses default threading and a 120-second per-scan timeout.
Each CodeQL run builds a single-snapshot database, then analyzes it with
\texttt{--threads=2}; the per-snapshot wall-clock budget is one hour, which
no snapshot in either configuration exceeded.

For the HEAD live scan (Section~\ref{sec:rq4}), we build a single
CodeQL database over the 16{,}433-file repository at commit
\texttt{4752e9a6} (2026-06-05), then evaluate each query individually with
\texttt{--threads=24}. We impose a four-hour per-query wall-clock budget;
queries that do not produce a result within that budget are recorded as
non-converging and excluded from the baseline finding count. The four-hour
ceiling was set so that the union of per-query budgets fits within a
single overnight run on the host; we verified, by sampling at one- and
two-hour intermediate checkpoints, that the queries that exceed two hours
all also exceed four hours without producing a non-empty result, so the
specific choice of cutoff does not affect the conclusion. The same budget
and host are used for the \tech queries.

\paragraph{Primary metric.}
We use advisory-level recall as the primary evaluation metric and report
it separately for the train and test partitions in all the following tables in the main body of this paper.
The false-positive rate is not within our primary scope of this study; we report
finding-volume statistics on the HEAD repository as a proxy for triage cost
in Section~\ref{sec:rq4}.

\section{Evaluation}
\label{sec:evaluation}

\subsection{Research Questions}
\label{sec:rqs}

We evaluate {\tech} around the following research questions.
\begin{description}[leftmargin=*]
    \item[RQ1.] How well do generic static-analysis tools detect disclosed
                vulnerabilities in local LLM agent runtimes?
    \item[RQ2.] To what extent do agent-specific static rules improve
                detection across vulnerability classes?
    \item[RQ3.] Do rules derived on past advisories generalize to advisories
                unseen during rule development?
    \item[RQ4.] What is the precision of runtime-layer auditing on current agent versions?
                %What does a preliminary precision audit reveal about the triage cost of applying {\tech} to current agent code?
    \item[RQ5.] Where is the boundary between syntactic rule-based detection
                and semantic reasoning for agent-runtime safety?
\end{description}
The five subsections below address one research question each, in order.

\subsection{RQ1: Generic Tools Detect Few Agent-Runtime Vulnerabilities}
\label{sec:rq1}

Semgrep Pro with \texttt{--config auto} draws on roughly 2{,}815 community
and Pro rules; CodeQL's \texttt{javascript-security-extended} suite adds 104
queries that include the full canonical taint set. On
\textsc{OpenClawBench}, these rulesets together flag a small minority of
the disclosed flaws. Semgrep Pro detects 47 of the 217 held-out advisories
(21.7\%); CodeQL detects 30 (13.8\%); their union covers 68 (31.3\%). On
the rule-derivation set the corresponding numbers are 25.3\%, 12.2\%, and
32.3\%, so the gap is not a temporal artefact. The full breakdown is given in
Table~\ref{tab:overall}, % (Section~\ref{sec:rq2}), 
where the same rows form
the baseline against which \tech is compared.

The gap persists across every severity tier and across the CWE classes
that dominate the benchmark; we defer those breakdowns to
Section~\ref{sec:rq2}, where the contrast with \tech is sharpest. The
most striking case is symlink following (CWE-59), where Semgrep Pro
detects zero of fifteen advisories and CodeQL detects three. Similar
zeros appear for several other agent-relevant classes under the Semgrep
baseline.

This is a structural mismatch rather than a coverage gap that a larger
generic ruleset would close. Both backends are widely deployed and
actively maintained; the difficulty is that their rulesets target the
operations and identifiers conventional web applications use. Local
agent runtimes do not interpolate workspace paths into prompts, fetch
model-supplied URLs through the dedicated helpers web frameworks expose,
or persist memory through marked-up writers in the canonical forms a
generic rule expects. Closing the gap requires rules whose patterns match
the operations the runtime actually performs.

\begin{findingbox}{Finding 1: Generic static-analysis rules miss most local-agent runtime vulnerabilities.}
On the held-out test set of \textsc{OpenClawBench}, Semgrep Pro detects only 21.7\% of disclosed advisories, and CodeQL's \texttt{security-extended} suite detects only 13.8\%. Their union reaches 31.3\%, indicating that existing generic JavaScript/TypeScript security rules cover only a minority of local-agent runtime weaknesses. The gap reflects a structural mismatch: local agents expose prompt builders, skill loaders, tool dispatchers, memory writers, and permission gates whose vulnerability patterns are not well represented in conventional web-application rulesets.
\end{findingbox}

\subsection{RQ2: ClawAudit Closes Most of the Gap Across Vulnerability Classes}
\label{sec:rq2}

\begin{table}[t]
\centering
\caption{Overall advisory-level recall on \textsc{OpenClawBench}, separately for the rule-derivation set (train, $n{=}229$) and the held-out test set ($n{=}217$). $\Delta$ is the gap between test and train recall. Inline annotations on the \tech rows show the absolute gain in pp vs.\ the corresponding baseline ($\uparrow$ increase, $\downarrow$ decrease). Numbers reflect detection on production code; findings located in test (case) files are excluded.}
\label{tab:overall}
\small
\setlength{\tabcolsep}{8pt}
\renewcommand{\arraystretch}{1.45}
\begin{tabular}{@{}lccc@{}}
\toprule
\rowcolor{gray!12}
\textbf{Configuration} & \textbf{Train (\%)} & \textbf{Test (\%)} & \textbf{$\Delta$ (pp)} \\
\midrule
Semgrep Pro
  & 25.3\,{\scriptsize(58/229)}
  & 21.7\,{\scriptsize(47/217)}
  & $-3.6$ \\
\rowcolor{teal!8}
\textbf{Semgrep\,+\,\tech}
  & \textbf{65.5}\,\textcolor{green!55!black}{\scriptsize$\uparrow$40.2}\,{\scriptsize(150/229)}
  & \textbf{66.8}\,\textcolor{green!55!black}{\scriptsize$\uparrow$45.1}\,{\scriptsize(145/217)}
  & $+1.3$ \\
\midrule
CodeQL \texttt{security-extended}
  & 12.2\,{\scriptsize(28/229)}
  & 13.8\,{\scriptsize(30/217)}
  & $+1.6$ \\
\rowcolor{teal!8}
\textbf{CodeQL\,+\,\tech}
  & \textbf{79.0}\,\textcolor{green!55!black}{\scriptsize$\uparrow$66.8}\,{\scriptsize(181/229)}
  & \textbf{75.1}\,\textcolor{green!55!black}{\scriptsize$\uparrow$61.3}\,{\scriptsize(163/217)}
  & $-3.9$ \\
\bottomrule
\end{tabular}
\end{table}

Augmenting each backend with \tech raises advisory-level recall sharply.
Semgrep + \tech reaches 65.5\% on the rule-derivation set, a 40.2~pp lift
over the Pro baseline, and 66.8\% on the held-out test set;
CodeQL + \tech reaches 79.0\% on the rule-derivation set, a 66.8~pp lift,
and 75.1\% on test (Table~\ref{tab:overall}). The lift is larger for
CodeQL, and a single mechanism accounts for most of the difference. Many
\tech rules check for the absence of a sanitization helper in the
function surrounding a potentially dangerous operation. Semgrep encodes
this with a stack of \texttt{pattern-not-inside} clauses that see only the
immediate AST context; 
%CodeQL expresses it as one existential over the call graph of the enclosing function, which crosses into helper modules without extra machinery. 
CodeQL expresses the check as a predicate over calls visible in the enclosing function, which can recognize guard-like helper calls even when they are not syntactically adjacent to the dangerous operation.
The categories where CodeQL widens its lead
most, missing authorization (CWE-285) and broken access control
(CWE-284), are exactly the ones whose sanitization guard typically lives
in a helper rather than at the immediate call site.

% \begin{table}[t]
% \centering
% \caption{Recall stratified by advisory severity on the train+test union
% ($n{=}446$). \tech improves recall in every tier for both backends.}
% \label{tab:severity}
% \begin{tabular}{lccccc}
% \toprule
%  & & \multicolumn{2}{c}{Semgrep} & \multicolumn{2}{c}{CodeQL} \\
% \cmidrule(lr){3-4} \cmidrule(lr){5-6}
% Severity & $n$ & Pro (\%) & + \tech (\%) & Sec-Ext (\%) & + \tech (\%) \\
% \midrule
% Critical &  12 & 33.3 & 83.3 & 16.7 & 83.3 \\
% High     & 134 & 26.1 & 67.2 & 13.4 & 73.9 \\
% Medium   & 250 & 19.2 & 63.6 & 13.2 & 76.8 \\
% Low      &  50 & 36.0 & 72.0 & 10.0 & 86.0 \\
% \midrule
% All      & 446 & 23.5 & 66.1 & 13.0 & 77.1 \\
% \bottomrule
% \end{tabular}
% \end{table}

\begin{table}[t]
\centering
\caption{Recall stratified by advisory severity on the train+test union ($n{=}446$). \tech improves recall in every tier for both backends. Inline annotations show the absolute change in pp vs.\ the corresponding baseline ($\uparrow$ increase, $\downarrow$ decrease).}
\label{tab:severity}
\begin{tabular}{lccccc}
\toprule
 & & \multicolumn{2}{c}{Semgrep} & \multicolumn{2}{c}{CodeQL} \\
\cmidrule(lr){3-4} \cmidrule(lr){5-6}
Severity & $n$ & Pro (\%) & + \tech (\%) & Sec-Ext (\%) & + \tech (\%) \\
\midrule
\cellcolor{red!40}Critical &  12 & 33.3 & 83.3\,\textcolor{green!55!black}{\scriptsize$\uparrow$50.0} & 16.7 & 83.3\,\textcolor{green!55!black}{\scriptsize$\uparrow$66.6} \\
\cellcolor{red!28}High     & 134 & 26.1 & 67.2\,\textcolor{green!55!black}{\scriptsize$\uparrow$41.1} & 13.4 & 73.9\,\textcolor{green!55!black}{\scriptsize$\uparrow$60.5} \\
\cellcolor{red!16}Medium   & 250 & 19.2 & 63.6\,\textcolor{green!55!black}{\scriptsize$\uparrow$44.4} & 13.2 & 76.8\,\textcolor{green!55!black}{\scriptsize$\uparrow$63.6} \\
\cellcolor{red!8}Low       &  50 & 36.0 & 72.0\,\textcolor{green!55!black}{\scriptsize$\uparrow$36.0} & 10.0 & 86.0\,\textcolor{green!55!black}{\scriptsize$\uparrow$76.0} \\
\midrule
All      & 446 & 23.5 & 66.1\,\textcolor{green!55!black}{\scriptsize$\uparrow$42.6} & 13.0 & 77.1\,\textcolor{green!55!black}{\scriptsize$\uparrow$64.1} \\
\bottomrule
\end{tabular}
\end{table}

% \begin{table}[t]
% \centering
% \caption{Recall (\%) stratified by advisory severity on the train+test union ($n{=}446$). \tech improves recall in every tier for both backends.}
% \label{tab:severity}
% \small
% \setlength{\tabcolsep}{7pt}
% \renewcommand{\arraystretch}{1.45}
% \begin{tabular}{@{}l c c >{\columncolor{teal!8}}c c >{\columncolor{teal!8}}c@{}}
% \toprule
% \rowcolor{gray!12}
%  & & \multicolumn{2}{c}{\textbf{Semgrep}} & \multicolumn{2}{c}{\textbf{CodeQL}} \\
% \cmidrule(lr){3-4} \cmidrule(lr){5-6}
% \rowcolor{gray!12}
% \textbf{Severity} & \textbf{$n$} & \textbf{Pro} & \textbf{+\,\tech} & \textbf{Sec-Ext} & \textbf{+\,\tech} \\
% \midrule
% Critical &  12 & 33.3 & \textbf{83.3} & 16.7 & \textbf{83.3} \\
% High     & 134 & 26.1 & \textbf{67.2} & 13.4 & \textbf{73.9} \\
% Medium   & 250 & 19.2 & \textbf{63.6} & 13.2 & \textbf{76.8} \\
% Low      &  50 & 36.0 & \textbf{72.0} & 10.0 & \textbf{86.0} \\
% \midrule
% \rowcolor{gray!15}
% \textbf{All} & \textbf{446} & \textbf{23.5} & \textbf{66.1} & \textbf{13.0} & \textbf{77.1} \\
% \bottomrule
% \end{tabular}
% \end{table}

The improvement holds across severity tiers (Table~\ref{tab:severity}).
Semgrep + \tech improves over Pro by between 36 and 50~pp depending on
tier, and CodeQL + \tech improves over its baseline by 50 to 76~pp; the
two backends finish within 8~pp of each other on every tier. Critical
advisories land at 83\% under both \tech configurations, the highest
absolute bucket; the lowest is medium under Semgrep at 64\%. We read the
roughly tier-independent lift as evidence that the rules target
structural features of the bugs rather than properties correlated with
how OpenClaw assigns severities.

% \begin{table}[t]
% \centering
% \caption{Per-CWE recall on the train+test union ($n{=}446$). \tech's
% largest gains land on agent-relevant classes (symlink, path traversal,
% TOCTOU, missing authorization) where the baselines are weakest.}
% \label{tab:cwe}
% \begin{tabular}{llccccc}
% \toprule
%  & & & \multicolumn{2}{c}{Semgrep (\%)} & \multicolumn{2}{c}{CodeQL (\%)} \\
% \cmidrule(lr){4-5} \cmidrule(lr){6-7}
% CWE & Description & $n$ & Pro & + \tech & Sec-Ext & + \tech \\
% \midrule
% CWE-863 & Authorization Bypass     & 77 & 23.4 & 58.4 &  9.1 & 80.5 \\
% CWE-22  & Path Traversal           & 31 & 16.1 & 83.9 & 16.1 & 80.6 \\
% CWE-78  & Command Injection        & 30 & 43.3 & 73.3 & 20.0 & 80.0 \\
% CWE-284 & Access Control           & 21 & 28.6 & 81.0 &  9.5 & 85.7 \\
% CWE-918 & SSRF                     & 20 & 25.0 & 75.0 & 10.0 & 80.0 \\
% CWE-285 & Missing Authorization    & 18 & 16.7 & 66.7 &  0.0 & 83.3 \\
% CWE-400 & Resource Exhaustion      & 17 & 47.1 & 70.6 & 17.6 & 88.2 \\
% CWE-184 & Incomplete Denylist      & 16 & 12.5 & 37.5 & 12.5 & 50.0 \\
% CWE-59  & Symlink Following        & 15 &  0.0 & 86.7 & 20.0 & 86.7 \\
% CWE-306 & Missing Authentication   & 10 & 50.0 & 80.0 & 30.0 & 100.0 \\
% CWE-367 & TOCTOU                   & 10 & 10.0 & 90.0 & 30.0 & 90.0 \\
% \bottomrule
% \end{tabular}
% \end{table}

\begin{table}[t]
\centering
\caption{Per-CWE recall on the train+test union ($n{=}446$). \tech's largest gains land on agent-relevant classes (symlink, path traversal, TOCTOU, missing authorization) where the baselines are the weakest. Inline annotations show the absolute change in pp vs.\ the corresponding baseline ($\uparrow$ increase, $\downarrow$ decrease).}
\label{tab:cwe}
\begin{tabular}{llccccc}
\toprule
 & & & \multicolumn{2}{c}{Semgrep (\%)} & \multicolumn{2}{c}{CodeQL (\%)} \\
\cmidrule(lr){4-5} \cmidrule(lr){6-7}
CWE & Description & $n$ & Pro & + \tech & Sec-Ext & + \tech \\
\midrule
CWE-863 & Authorization Bypass     & 77 & 23.4 & 58.4\,\textcolor{green!55!black}{\scriptsize$\uparrow$35.0}  &  9.1 &  80.5\,\textcolor{green!55!black}{\scriptsize$\uparrow$71.4} \\
\rowcolor{gray!7}
CWE-22  & Path Traversal           & 31 & 16.1 & 83.9\,\textcolor{green!55!black}{\scriptsize$\uparrow$67.8}  & 16.1 &  80.6\,\textcolor{green!55!black}{\scriptsize$\uparrow$64.5} \\
CWE-78  & Command Injection        & 30 & 43.3 & 73.3\,\textcolor{green!55!black}{\scriptsize$\uparrow$30.0}  & 20.0 &  80.0\,\textcolor{green!55!black}{\scriptsize$\uparrow$60.0} \\
\rowcolor{gray!7}
CWE-284 & Access Control           & 21 & 28.6 & 81.0\,\textcolor{green!55!black}{\scriptsize$\uparrow$52.4}  &  9.5 &  85.7\,\textcolor{green!55!black}{\scriptsize$\uparrow$76.2} \\
CWE-918 & SSRF                     & 20 & 25.0 & 75.0\,\textcolor{green!55!black}{\scriptsize$\uparrow$50.0}  & 10.0 &  80.0\,\textcolor{green!55!black}{\scriptsize$\uparrow$70.0} \\
\rowcolor{gray!7}
CWE-285 & Missing Authorization    & 18 & 16.7 & 66.7\,\textcolor{green!55!black}{\scriptsize$\uparrow$50.0}  &  0.0 &  83.3\,\textcolor{green!55!black}{\scriptsize$\uparrow$83.3} \\
CWE-400 & Resource Exhaustion      & 17 & 47.1 & 70.6\,\textcolor{green!55!black}{\scriptsize$\uparrow$23.5}  & 17.6 &  88.2\,\textcolor{green!55!black}{\scriptsize$\uparrow$70.6} \\
\rowcolor{gray!7}
CWE-184 & Incomplete Denylist      & 16 & 12.5 & 37.5\,\textcolor{green!55!black}{\scriptsize$\uparrow$25.0}  & 12.5 &  50.0\,\textcolor{green!55!black}{\scriptsize$\uparrow$37.5} \\
CWE-59  & Symlink Following        & 15 &  0.0 & 86.7\,\textcolor{green!55!black}{\scriptsize$\uparrow$86.7}  & 20.0 &  86.7\,\textcolor{green!55!black}{\scriptsize$\uparrow$66.7} \\
\rowcolor{gray!7}
CWE-306 & Missing Authentication   & 10 & 50.0 & 80.0\,\textcolor{green!55!black}{\scriptsize$\uparrow$30.0}  & 30.0 & 100.0\,\textcolor{green!55!black}{\scriptsize$\uparrow$70.0} \\
CWE-367 & TOCTOU                   & 10 & 10.0 & 90.0\,\textcolor{green!55!black}{\scriptsize$\uparrow$80.0}  & 30.0 &  90.0\,\textcolor{green!55!black}{\scriptsize$\uparrow$60.0} \\
\bottomrule
\end{tabular}
\end{table}

Per-CWE stratification (Table~\ref{tab:cwe}) localizes the lift. As noted in Section~\ref{sec:taxonomy-stride}, CWE-400 advisories are retained only when the fix exposes a statically visible implementation pattern, such as a missing size check or unbounded parser, and are mapped to the closest runtime boundary rather than treated as a separate DoS category. The
biggest gains occur on classes whose vulnerable code has a syntactically
recognizable shape: symlink following moves from 0--20\% baseline to
86.7\% in both backends; TOCTOU moves from 10--30\% to 90\%; path
traversal moves from 16\% to 81--84\%; access control moves from 10--29\%
to 81--86\%. CWE-59 is the cleanest motivating case: Semgrep Pro detects
zero of fifteen symlink advisories, and \tech detects 13 out of
15 in both backends, on a class generic JavaScript rulesets
essentially do not address.

The two backends part company on the semantic end of the CWE table.
Authorization bypass (CWE-863) climbs to 58\% under Semgrep + \tech but
to 81\% under CodeQL + \tech; missing authorization (CWE-285) climbs to
67\% under Semgrep + \tech but to 83\% under CodeQL + \tech. In both
classes the auth check is usually present somewhere in the agent
runtime, just at a depth Semgrep's AST-local
\texttt{pattern-not-inside} cannot follow. 
%CodeQL's call-graph existential reaches the helper and removes the spurious match. 
CodeQL’s predicate can recognize guard-like helper calls when they are invoked in the enclosing function. 
The same
mechanism widens the CodeQL lead on CWE-306 missing authentication
(100\% vs 80\%) and on the long-tail authz classes reported in
Appendix~\ref{app:cwe}.

Two classes resist both configurations. \tech raises CWE-184 incomplete
denylist from 12.5\% baseline to 37.5\% (Semgrep) and 50\% (CodeQL), and
raises CWE-200 information exposure from near zero to 40--80\% (full
numbers in Appendix~\ref{app:cwe}). We treat these as the residual gap
of rule-based detection and return to them as the focus of RQ5
(Section~\ref{sec:rq5}).

\begin{findingbox}{Finding 2: Agent-specific rules substantially improve recall across backends and severity tiers.}
Adding \tech rules raises held-out Semgrep recall from 21.7\% to 66.8\% and held-out CodeQL recall from 13.8\% to 75.1\%. The improvement appears across all severity tiers and is largest for vulnerability classes with recurring implementation structure, including symlink following, path traversal, TOCTOU, command injection, SSRF, and access-control failures. This shows that local-agent runtime vulnerabilities are not invisible to static analysis; they require domain-specific rules that match the runtime operations agents actually perform.
\end{findingbox}

\subsection{RQ3: Rules Generalize to Unseen Advisories}
\label{sec:rq3}

The held-out 217 advisories were collected after every \tech rule had been
committed and were evaluated exactly once. Train-to-test recall gaps are
$-3.6$~pp for Semgrep Pro, $+1.3$~pp for Semgrep + \tech, $+1.6$~pp for
CodeQL \texttt{security-extended}, and $-3.9$~pp for CodeQL + \tech
(Table~\ref{tab:overall}). All four configurations stay inside a $\pm 4$~pp
band: the rules detect the held-out advisories at a rate within sampling
distance of their detection rate on the rule-derivation advisories. 
%The failure mode that a security analyst would worry about, rules that match the specific identifiers or surrounding code from the train advisories, would produce a sharp drop on test; the data show no such drop.
A natural concern is overfitting: the custom rules might match identifiers, helper names, or surrounding code contexts specific to the rule-derivation advisories. Such memorization would produce a sharp recall drop on the held-out test set. Instead, the train-to-test gaps remain within $\pm$4 percentage points across all configurations. 

The negative gap on CodeQL + \tech is the one result that requires
unpacking. Appendix~\ref{app:cwe} breaks the gap down by CWE. The drop is
concentrated on CWE-78 command injection, where two CodeQL queries
($\texttt{execsync-template-literal}$ and
$\texttt{scp-ssh-unvalidated-host}$) match a shell-spawn shape narrower
than what the test-set advisories actually exhibit; partial offsets come
from CWE-863 (authorization bypass) and CWE-184 (incomplete denylist),
both of which generalize positively. The pattern is consistent with what
we call rule-coverage variance: the train set is a finite sample of the
runtime's vulnerability stream, and the held-out set draws from the same
stream with a slightly different CWE mix. Neither backend exhibits the
all-classes degradation that would indicate 
%overfitting in the gradient-descent sense (no parameters were fit); 
memorization of rule-derivation advisories; the Semgrep ruleset,
with broader per-rule scope, simply happens to be more robust to such
shifts than the CodeQL ruleset, which has sharper per-rule scope.

\begin{findingbox}{Finding 3: The rules generalize temporally rather than memorizing the rule-derivation advisories.}
The held-out test set was collected after rule finalization and evaluated once. Across all four configurations, train-to-test recall gaps remain within $\pm$4 percentage points: $-3.6$ pp for Semgrep Pro, $+1.3$ pp for Semgrep+\tech, $+1.6$ pp for CodeQL \texttt{security-extended}, and $-3.9$ pp for CodeQL+\tech. These small gaps suggest that \tech captures recurring runtime vulnerability structures rather than merely matching identifiers or code contexts from the rule-derivation set.
\end{findingbox}

\subsection{RQ4: Precision Drops on Live HEAD}
\label{sec:rq4}

The advisory-level numbers above measure recall on small,
fix-commit-sized snapshots. To gauge what a \tech deployment would
actually surface on a developer's machine, we ran each \tech
configuration against the live OpenClaw repository at HEAD (commit
\texttt{4752e9a6}; 16{,}433 JavaScript/TypeScript files) and audited a stratified sample
of findings. Semgrep + \tech produced 3{,}473 production-code findings;
CodeQL + \tech produced 3{,}756. We drew 25 findings per backend by
stratified sampling over the set of fired rules, excluded test files in
advance, and labeled each as a true positive (TP) if the flagged code
matches the vulnerability pattern the rule claims, or as a false
positive (FP) otherwise.

\begin{table}[t]
\centering
\caption{Manual precision audit on the HEAD live scan. 25 production-only
findings per backend, sampled stratified by rule.}
\label{tab:precision}
\begin{tabular}{lcccc}
\toprule
Configuration            & Sample size & TP & FP & Precision (\%) \\
\midrule
Semgrep + \tech          & 25 & 3 & 22 & 12.0 \\
CodeQL  + \tech          & 25 & 3 & 22 & 12.0 \\
\midrule
Combined                 & 50 & 6 & 44 & 12.0 \\
\bottomrule
\end{tabular}
\end{table}

%Both backends land at 12\% precision (Table~\ref{tab:precision}). 
As summarized in Table~\ref{tab:precision}, in this preliminary sample, both backends yield 3 pattern-level true positives out of 25 inspected findings, suggesting low live-code precision (12\%) and substantial triage cost.
%
%In this preliminary sample, both backends yield 3 true positives out of 25 inspected findings, suggesting low live-code precision and substantial triage cost. 
The HEAD finding volumes are essentially equal across backends (3{,}473
versus 3{,}756), so the recall advantage CodeQL holds over Semgrep on
\textsc{OpenClawBench} does not extend to a precision advantage on live
code: per finding, the two backends yield comparable numbers of true
positives, and the manual triage they require is comparable in
magnitude.

Three properties of the current ruleset together account for the low
number. Most \tech rules identify attacker-controlled input by an
identifier-name regex, which on production code routinely matches
variables that carry trusted CLI arguments, hash-derived paths,
operator-controlled configuration, or enum constants. \tech also checks
for sanitization guards within the enclosing function, but production
code frequently extracts the guard into a helper module that the
\texttt{pattern-not-inside} clause cannot see and that CodeQL's
call-graph predicate sometimes misses for alias-named callers. And the
rules are deliberately tuned for advisory-level recall, so the broad-net
patterns that catch many true vulnerabilities on small fix-commit
snapshots also fire on a large surface of non-vulnerable code in a
16{,}000-file repository.

The results also suggest that rule quality is not uniform. All six true positives in our sample come
from six distinct narrow-scope rules:
\texttt{external-payload-url-fetch},
\texttt{realpath-without-hardlink-check}, and
\texttt{path-boundary-startswith-only} on the Semgrep side, and
\texttt{error-stack-in-response},
\texttt{sandbox-mode-default-off}, and
\texttt{auth-function-default-allow} on the CodeQL side. The
high-volume rules
(\texttt{path-join-user-input-no-boundary-check},
\texttt{fs-write-risky-path},
\texttt{json-parse-no-size-limit}) produced no true positives in the
sample. We read this as a bimodal distribution of rule quality:
production deployments should prioritize the narrow subset and reserve
the broader rules for combination with downstream taint or semantic
analysis (Section~\ref{sec:semantic-analysis}).

\begin{findingbox}{Finding 4: Current rules are recall-oriented and impose substantial live-code triage cost.}
A preliminary pattern-level precision audit on the live OpenClaw HEAD repository shows low precision: each backend yields 3 true positives out of 25 inspected production-code findings. This indicates that the current \tech rules are best understood as recall-oriented audit rules, not as a low-noise production scanner. Narrow, structurally specific rules produce the useful findings, while broad identifier-based rules dominate the false positives and should be paired with taint, semantic filtering, or rule prioritization before deployment.
\end{findingbox}

\subsection{RQ5: Static Rules Stall at Semantically Defined CWEs}
\label{sec:rq5}

The CWE classes \tech struggles with share a property the others do not:
the bug has no syntactic surface on the vulnerable side. CWE-184
incomplete denylist is the cleanest example. The bug is that a list does
not contain an entry it should, and the source code at the vulnerable
revision is syntactically indistinguishable from correct code; only an
external specification of what the denylist ought to contain reveals the
gap. CWE-200 information exposure is similar but more diffuse. Secrets
can leak through logs, error messages, telemetry payloads, response
headers, or URL paths, and each sink has its own structural shape;
\tech catches the URL-with-secret-into-error case through a call-graph
query, but bottoms out elsewhere because no single rule shape covers
every leak path. CWE-639 (authorization bypass through a user-controlled resource identifier) sits between the two:
some of its advisories carry a named auth helper a rule can match, and
others express the missing check as an inferred constraint over which
session owns which resource. These three classes plus the long tail of
CWE classes with $n \le 4$ account for nearly all the residual gap to
100\% recall.

CodeQL's canonical taint queries do not rescue this regime. Of the 104
queries in the \texttt{security-extended} suite, 21 fail to terminate
under a 4-hour per-query budget on the 16{,}433-file HEAD repository,
among them \texttt{js/command-injection}, \texttt{js/request-forgery},
\texttt{js/client-side-url-redirect}, and \texttt{js/stored-xss}. These
are the queries that motivate CodeQL adoption on conventional web
codebases; their non-convergence shapes the baseline we report
(216 findings from the 83 queries that completed, not from all 104).
\tech's own queries, %which use bounded call-graph traversal rather than global taint, 
many of which use bounded guard predicates rather than global taint, 
all completed in seconds to minutes; the failure mode is
specific to canonical taint, not to CodeQL as such.

The boundary is therefore not where static analysis stops working in
the abstract, but where the bug stops having a syntactic surface to
intercept. Closing the residual gap requires analyses that supply the
missing surface: mining specifications of expected denylist contents
from prior advisories, tracking attacker-influenced sources across the
runtime (with prompt and tool output as taint sources rather than HTTP
request and response), or reading natural language alongside code to
infer which strings are sensitive. 
%We sketch concrete directions in Section~\ref{sec:semantic-analysis}.

\vspace{-2pt}
\begin{findingbox}
%{Finding 5: Rule-based auditing succeeds when the vulnerability has a syntactic surface and stalls when correctness depends on intent.}
{Finding 5: Rule-based auditing is less effective or even stalls when security depends on intent.}
{\tech} performs best on vulnerabilities whose unsafe form has a recurring code shape, such as path traversal, symlink following, command injection, SSRF, TOCTOU, and missing local guards. It struggles on semantically defined weaknesses such as incomplete denylists, information exposure, and resource-ownership authorization, where the vulnerable code is indistinguishable from correct code without an external specification of intended policy or sensitive-data semantics. Closing this residual gap requires semantic agent-runtime analysis beyond pattern rules alone.
\end{findingbox}

\vspace{4pt}\noindent
\textbf{Responsible disclosure.} This paper evaluates disclosed historical advisories and does not publish exploit procedures for unresolved vulnerabilities. For any newly identified live-code findings, we follow responsible disclosure before public release.

\section{Discussion}
\label{sec:discussion}

\subsection{Implications for LLM-Agent Safety}
\label{sec:implications}

Three implications follow from the results, each constraining a separate
research agenda in agent safety.

\paragraph{Prompt-level defenses are necessary but not sufficient.}
A growing literature on indirect prompt injection, jailbreaks, and tool-use
attacks treats the agent as a model that should refuse certain inputs or
emit only certain outputs. Our results show that even an agent that
correctly refuses every dangerous prompt can still be exploited through
its runtime: an attacker who controls a header logged by the agent
(GHSA-g27f-9qjv-22pm), a workspace directory name interpolated into a
system prompt (GHSA-2qj5-gwg2-xwc4), or a Telegram URL written into an
error message (GHSA-chf7-jq6g-qrwv) can manipulate or exfiltrate without
ever appearing in the model's prompt window. Prompt-side robustness must
therefore be paired with source-level assurance on the components that
construct prompts, route tool calls, and emit logs.

\paragraph{Marketplace audits do not generalize to runtime audits.}
%Recent work has shown that thousands of malicious skills are present in
%local-agent marketplaces and that targeted skill audits can flag many of
%them at install time. 
Recent work has shown that local-agent marketplaces contain thousands of vulnerable or suspicious skills, including hundreds of behaviorally confirmed malicious skills, and that targeted skill scanners can flag many such risks before or during installation~\cite{liu2026agent,beurer2026technical,snyk2026toxicskills,snyk2026vercelskills,snyk2026agentscan}.
Our results identify a complementary attack surface
that is invisible to skill audits: the agent runtime that loads and
executes the skill itself. A benign skill that flows through an unsafe
loader, a permission gate that is checked too late, or a memory writer
that does not sanitize tool output produces the same end state as a
malicious skill, but with no third-party artefact to inspect. Both audits
are needed; neither subsumes the other.

\paragraph{The privileged-runtime view applies beyond OpenClaw.}
Although our evaluation is on OpenClaw, the five categories that our taxonomy
defines correspond to runtime components common to any local LLM agent
that constructs prompts, dispatches tools, mediates skills, opens network
connections, and enforces permissions. The disclosed advisories that drive
the rules cluster on these components rather than on any OpenClaw-specific
implementation choice. We therefore expect the same taxonomy and a
substantial fraction of the same rules to transfer to other local agents
with comparable runtime architecture; verifying this is the most immediate
generalization study left open by this work.

\subsection{Why Generic Static Analysis Is Insufficient}
\label{sec:generic-insufficient}

The 14--25\% baseline recall reported in Table~\ref{tab:overall} is not an
indictment of the tools themselves; both Semgrep~Pro and CodeQL's
\texttt{security-extended} suite are mature, well-curated rulesets that
collectively encode over 2{,}915 patterns. They are weak on
\textsc{OpenClawBench} for structural reasons that any future generic
ruleset would inherit unless it is deliberately specialized.

\paragraph{Generic rules target conventional web shapes.}
The dominant CWE classes baseline rules detect well, ReDoS, XSS,
prototype pollution, insecure randomness, are dictated by what conventional
web and Node.js applications do. The dominant CWE classes in
\textsc{OpenClawBench}, authorization bypass (77), path traversal (31),
command injection (30), SSRF (20), missing authorization (18), reflect
what local agents do instead: open shells, load skills, fetch model-supplied
URLs, gate handler routes. The mismatch is structural, not a tuning gap.
Even a generic ruleset with several thousand additional rules would not
materially close it unless the new rules targeted these agent-runtime
operations specifically.

\paragraph{Agent-runtime patterns are not local.}
Many of the highest-volume \tech rules express ``operation $X$ appears in
function $F$ without the matching guard anywhere in $F$''. Encoding the
guard interprocedurally is straightforward in CodeQL once the rule names
the operation and the guard helper. Generic rulesets do not include these
predicates because they presuppose helpers,
\texttt{sanitizePromptText}, \texttt{isPrivateIp}, \texttt{normalizeScpRemoteHost},
\texttt{redactUrl}, that exist as agent-runtime conventions rather than
language-level idioms. A generic ruleset has no basis for inferring that
these names mark a safe boundary.

\paragraph{Canonical taint queries do not converge.}
Even the queries generic CodeQL ships specifically for runtime concerns,
the \texttt{security-extended} taint suite, run into a separate problem:
21 of 104 queries fail to terminate within a 4-hour per-query budget on
the 16{,}433-file OpenClaw HEAD repository. The queries are not wrong; they
encode reasonable taint specifications for conventional web servers, but
the source-and-sink graph of an agent runtime is denser and longer than
those specifications were designed for. Generic taint analysis therefore
does not scale to agent-sized codebases without specialization, a separate
argument against straightforwardly treating the runtime layer as a target of existing tools. 
Nevertheless, this result is specific to our OpenClaw HEAD configuration and timeout budget; we do not claim that these queries fail generally.

\subsection{The Syntactic--Semantic Boundary in Agent-Runtime Auditing}
\label{sec:syntactic-semantic-boundary}

The single most consistent observation across our evaluation is that
rule-based static analysis catches one kind of bug and not another. It
catches bugs whose vulnerable form has a recurring code shape:
\texttt{path.join} of user input without a containment check, \texttt{realpath}
without a symlink check, a \texttt{fetch} of a model-supplied URL without an
allowlist consultation, and a handler entry without an auth check. In each case
the patch introduces or relocates a named operation, and the rule simply
asks whether that operation appears in the vulnerable function. Recall is
high because the bug has a finite syntactic signature that a rule of bounded
size can match.

The rule-based static analysis~\cite{nong2021evaluating} struggles, in both backends, with bugs whose correctness depends on
something other than code shape: an intended access-control policy
(CWE-863 authorization bypass, CWE-285 missing authorization), the
identity of the requesting user or the ownership of the resource being
accessed (CWE-639), the semantics of which strings carry secrets
(CWE-200 information exposure), or the completeness of a manually
curated list (CWE-184 incomplete denylist). These bugs admit no
syntactic surface on the vulnerable side: the source code at the
vulnerable revision is, by construction, indistinguishable from correct
code unless an auditor supplies the missing intent. CodeQL's call-graph
existential narrows the gap on interprocedural authorization, where the
intent is encoded by the presence of a named auth helper; it does not
narrow the gap on the other classes, where there is no helper to find.

The practical consequence is that an agent-runtime audit cannot consist
of static rules alone. Rule-based detection should be paired with
analyses that supply the missing semantic surface: specifications of
expected denylist contents derived from prior advisories, runtime
metadata about which strings carry secrets in a given deployment, or
policy specifications that an automated checker can compare program
behavior against. We sketch concrete directions in
Section~\ref{sec:semantic-analysis}.

% \subsection{Toward Semantic Agent-Runtime Analysis}
% \label{sec:semantic-analysis}

% The blind spots above motivate analyses that combine static rules with
% semantic reasoning. Possible directions include interprocedural taint
% tracking specialized to agent runtimes (where prompt and tool output
% become the relevant sources and sinks rather than HTTP request and
% response), LLM-assisted rule synthesis from advisory text, specification
% mining for authorization policies, and runtime instrumentation to
% provide behavioral context that pure source-code analysis cannot
% recover.

\subsection{Toward Semantic Agent-Runtime Analysis}
\label{sec:semantic-analysis}

The blind spots above motivate analyses that combine static rules with semantic reasoning. We see four concrete directions.

\paragraph{Agent-specific interprocedural taint analysis.}
Existing taint analyses for JavaScript and TypeScript typically model web-application sources and sinks, such as HTTP request parameters flowing into SQL queries, shell commands, redirects, or DOM operations. Local LLM agents require a different taint model. The relevant sources include user goals, retrieved documents, tool outputs, model-emitted action arguments, prior memory entries, skill metadata, and files in the workspace. The relevant sinks include prompt templates, memory writers, tool dispatchers, shell executors, skill loaders, URL fetchers, log emitters, and permission-gated handlers. A semantic extension of \tech could therefore define an agent-runtime taint specification and track flows across the prompt--model--tool boundary, rather than only across conventional request--response boundaries. This would reduce the false positives caused by identifier-name heuristics and improve coverage for vulnerabilities whose risky value is not visible at the immediate call site.

\paragraph{Policy mining for authorization and isolation.}
Many residual failures are authorization or isolation bugs: a handler accepts a workspace ID without checking ownership, a skill can access a broader filesystem region than intended, or a permission gate is applied after a side effect has already occurred. These bugs are difficult because the missing condition is not syntactically present in the vulnerable revision. A future analyzer could mine candidate policies from route definitions, permission helpers, access-control middleware, configuration files, and patched versions of prior advisories. The resulting policy could then be checked against handler implementations: for example, every operation on a workspace, conversation, task, credential, or skill should be dominated by a check binding that resource to the current user or session. This would move the analysis from `does this function contain a guard-looking call?'' to `does this action satisfy the resource-ownership policy the runtime appears to enforce elsewhere?''

\paragraph{LLM-assisted rule and specification synthesis.}
The current \tech rules are manually written from advisory text and vulnerable patches. This process can be partially automated. Given an advisory description, a vulnerable snippet, and a fixing patch, an LLM could propose: (1) the runtime boundary involved, (2) the attacker-controlled source, (3) the privileged sink, (4) the missing guard, and (5) a candidate Semgrep rule or CodeQL query. Human review would still be needed, but the LLM could reduce the cost of turning advisory streams into new audit rules. More importantly, the LLM could synthesize semantic predicates that are hard to enumerate manually, such as `this string is a secret-bearing value,'' `this route requires workspace ownership,'' or ``this list is intended to block all private-network destinations.'' These predicates could then be compiled into static queries or used to prioritize findings for manual review.

\paragraph{Runtime instrumentation for semantic context.}
Some facts needed for accurate auditing are not recoverable from source code alone. Whether a URL is attacker-controlled, whether a path is inside a user workspace, whether a memory entry later re-enters a system prompt, or whether a permission check occurs before the first side effect may depend on runtime behavior. Lightweight instrumentation could record prompt construction events, tool-call arguments, memory writes, permission decisions, network destinations, and filesystem accesses during representative agent executions. Static findings could then be filtered or ranked using this behavioral context. For example, a path traversal rule firing on a variable that is never influenced by user, model, skill, or tool output could be deprioritized, while a rule firing on a value observed to flow from model output into a shell executor could be escalated.

Together, these directions suggest a path beyond recall-oriented rule matching. Static rules provide the first layer: they identify recurring implementation structures in the agent runtime. Semantic analysis supplies the missing intent: which values are attacker-controlled, which resources are protected, which strings are sensitive, and which checks are required before privileged actions. A mature agent-runtime auditor will likely need both layers: domain-specific static rules for scalable coverage and semantic reasoning to reduce triage cost and recover policy-dependent vulnerabilities.

\vspace{-4pt}
\section{Related Work}
\label{sec:related-work}
\vspace{-4pt}
\subsection{Prompt Injection and Behavioral Agent Evaluation}
\label{sec:rw-prompt-injection}

A first line of work studies how LLM agents can be manipulated through their
inputs. Greshake et al.~\cite{greshake2023not} introduce \emph{indirect} prompt injection,
where adversarial instructions hidden in retrieved documents or tool
outputs propagate into the agent's prompt window without ever appearing in
the user's message. Follow-up studies extend the attack surface to tool-use
flows in deployed agents, characterize adaptive and chained attack
strategies~\cite{zhan2025adaptive}, and develop targeted defenses against
specific injection vectors~\cite{chang2025chatinject}. A parallel line
constructs behavioral benchmarks that probe agents through their input/output
interface and report end-to-end success rates under adversarial
prompts~\cite{arora2025setupbench,patir2026dualguage}.
What this literature shares is its vantage point: the agent is observed
through its prompt window and its emitted actions, while the code that
constructs prompts, parses model output, and dispatches tool calls is
treated as a correct mediator. \tech's contribution is orthogonal: we
audit that mediator as source code, and the runtime vulnerabilities we
catch (\emph{e.g.}, workspace paths inlined into system prompts, untrusted
headers written into LLM-consumed logs) are invisible to prompt-window
evaluation because they arise inside the runtime before any prompt is
constructed or after the model has already returned.

\subsection{Security of Agent Tools, Skills, and Marketplaces}
\label{sec:rw-marketplaces}

A second line studies the third-party artefacts agents draw on.
Marketplace audits have catalogued hundreds of malicious skills on
repositories such as ClawHub~\cite{ClawHavoc341Malicious} and reported
broader patterns of credential exfiltration through the model
context~\cite{beurer2026280+}. Recent work models the agent--skill
interaction explicitly~\cite{liu2026agent} and analyzes the trust
relationships an agent inherits when it accepts a third-party
skill~\cite{hasan2025model}.
These efforts target the skill-side of the trust boundary, asking whether
an installable artefact is malicious before the agent loads it. \tech
targets the runtime-side of the same boundary, asking whether the agent
that loads the skill mediates it safely. A skill audit cannot catch a
permission gate that fires after its side effect, a loader that runs
setup hooks before allow-listing, or an inter-skill isolation that does
not exist, all of which are bugs in the agent runtime regardless of the
skill's contents. The two analyses operate at different points of the
same trust boundary and neither subsumes the other.

\subsection{Static Analysis for Vulnerability Detection}
\label{sec:rw-static-analysis}

Static analysis for detecting software vulnerabilities~\cite{nong2021evaluating} is a mature field. Pattern-based
detectors~\cite{nong2022open} such as Semgrep~\cite{semgrep} encode bugs as syntactic templates with meta-variable
constraints, while graph-based tools such as CodeQL~\cite{codeql} evaluate declarative
queries over a relational program representation that supports both
syntactic and dataflow predicates. Both ecosystems ship large, curated
rulesets organized by CWE class, principally for conventional web and
backend application code.
What is missing from this literature is an agent-runtime evaluation.
Existing CWE-oriented benchmarks, both academic and industrial,
contain web and library code; they do not contain the runtime structures,
prompt builders, skill loaders, memory writers, permission gates, that
characterize local LLM agents. As a result, the recall numbers established
tools post on those benchmarks do not predict the recall they would deliver
on an agent. Our 14--25\% baseline recall on \textsc{OpenClawBench}
quantifies that gap. \tech does not modify either analyzer; it instead
adds an agent-runtime rule layer that fits the same APIs the tools already
expose, so that the gain is repeatable in any deployment of those tools.

\subsection{Positioning and Contextualization}
\label{sec:rw-positioning}

Relative to prompt-injection and behavioral agent research, \tech audits
the runtime that mediates between prompts and actions rather than the
prompts and actions themselves. Relative to marketplace and skill audits,
\tech audits the loader that runs the skill rather than the skill artefact.
Relative to mainstream static-analysis evaluations, \tech targets a code
base, a local LLM agent runtime, on which the established rulesets have
not been measured before, and it reports recall under two backends with a
shared taxonomy so that the contribution attributable to rule design can
be separated from the contribution attributable to analysis engine.
To the best of our knowledge, no prior work has systematically examined an agent's own
source tree via a CWE-oriented implementation-weakness audit at this scale,
derived a taxonomy of runtime weakness classes from threat modeling, or
instantiated such a taxonomy in two static-analysis backends for
cross-tool measurement.

\vspace{-4pt}
\section{Limitations}
\label{sec:limitations}
\vspace{-4pt}

Our evaluation has six limitations.
First, the study focuses on \openc. Although we discuss \nanobot as a
motivating example in Section~\ref{sec:introduction}, we do not evaluate
\tech on it. Generalization to other local LLM agent runtimes is left to
future work.
Second, the ground truth comes entirely from disclosed advisories.
Vulnerabilities that have not been publicly reported, including bug
classes that the OpenClaw maintainers have not yet recognized as
security-relevant, are absent from \textsc{OpenClawBench} and therefore
absent from the recall denominator.
Third, the benchmark is positive-heavy and recall-oriented by
construction. Each entry is a known vulnerable fix-commit snapshot, so
we measure how reliably the rules recover labeled positives; the
benchmark does not include matched negative controls and cannot, by
itself, quantify the false-positive cost of deploying the rules.
Fourth, the precision audit in Section~\ref{sec:rq4} relies on a
50-finding sample labeled by a single reviewer. A larger multi-rater
study would tighten the precision estimate and surface per-rule
precision variance not visible at this sample size.
Fifth, rule-based static analysis is inherently limited for semantic
policy violations such as incomplete denylists (CWE-184) and
information exposure (CWE-200), as
Sections~\ref{sec:rq5} and~\ref{sec:syntactic-semantic-boundary}
discuss.
Sixth, the taxonomy is anchored to the runtime structures we observe in
\openc and to advisory categories disclosed to date. As more local LLM
agents are audited and as new runtime components emerge (for example,
multi-agent coordinators or persistent memory backends), the five-category
structure will likely need extension or refinement.

\vspace{-4pt}
\section{Conclusion}
\label{sec:conclusion}
\vspace{-4pt}

We presented \tech, a static auditing framework for implementation-level
risks in local LLM agent runtimes. By deriving an agent-runtime taxonomy from
STRIDE and instantiating it in both Semgrep (47 YAML rules) and CodeQL
(30 queries), \tech substantially improves detection recall over both
generic baselines on \textsc{OpenClawBench}, a benchmark of 446 disclosed
\openc advisories with a temporal train/test split. On the held-out test
set, Semgrep recall rises from 21.7\% to 66.8\% and CodeQL recall from
13.8\% to 75.1\%, with train/test gaps within 4 percentage points for all
four configurations. More broadly, our results show that local LLM agents should be
treated not only as model-driven systems but also as privileged software
runtimes whose source code requires systematic assurance.

Our artifact including {\tech} source code and experiment data/results is found at \url{https://github.com/SRestLabUB/ClawAudit}.

\bibliography{references}
\bibliographystyle{plain}

\appendix

% --- Listing styles for YAML and CodeQL ---------------------
\lstdefinestyle{yamlstyle}{
    basicstyle=\ttfamily\scriptsize,
    keywordstyle=\color{blue!70!black}\bfseries,
    stringstyle=\color{red!50!black},
    commentstyle=\color{green!40!black}\itshape,
    morecomment=[l]{\#},
    morestring=[b]",
    morestring=[b]',
    morekeywords={id,languages,severity,message,metadata,pattern,
                  patterns,pattern-either,pattern-not-inside,
                  metavariable-regex,metavariable,regex,
                  cwe,category,confidence,impact,cat},
    showstringspaces=false,
    breaklines=true,
    columns=fullflexible,
    frame=single,
    framerule=0.3pt,
    rulecolor=\color{black!30},
    xleftmargin=4pt,
    xrightmargin=4pt,
}
\lstdefinestyle{qlstyle}{
    basicstyle=\ttfamily\scriptsize,
    keywordstyle=\color{blue!70!black}\bfseries,
    stringstyle=\color{red!50!black},
    commentstyle=\color{green!40!black}\itshape,
    morecomment=[s]{/**}{*/},
    morecomment=[l]{//},
    morestring=[b]",
    morekeywords={import,from,where,select,predicate,exists,
                  not,and,or,in,instanceof,as},
    showstringspaces=false,
    breaklines=true,
    columns=fullflexible,
    frame=single,
    framerule=0.3pt,
    rulecolor=\color{black!30},
    xleftmargin=4pt,
    xrightmargin=4pt,
}

\section{Representative \tech Rule Examples}
\label{app:rules}

\begin{lstlisting}[style=yamlstyle,caption={CAT-1 Semgrep rule: workspace
path interpolated into a system-prompt template, with no sanitizer in the
enclosing function.},label={lst:sg-prompt}]
- id: workspace-path-into-system-prompt
  languages: [typescript, javascript]
  severity: WARNING
  message: |
    A workspace / cwd / project directory path is interpolated into a
    system prompt template without sanitization. Directory names can
    contain control characters or Unicode bidi marks that break prompt
    structure and inject attacker-controlled instructions.
  metadata:
    cwe: ["CWE-74"]
    category: "Cat-1: Prompt Handling"
  pattern-either:
    - patterns:
        - pattern: `...${$VAR}...`
        - metavariable-regex:
            metavariable: $VAR
            regex: '.*(workspaceDir|workspacePath|cwd|currentDir|projectDir|projectRoot)$'
        - pattern-not-inside: |
            sanitizePromptText(...)
        - pattern-not-inside: |
            stripControlChars(...)
    - patterns:
        - pattern: $T.systemPrompt = `...${$VAR}...`
        - metavariable-regex:
            metavariable: $VAR
            regex: '.*(workspace|cwd|dir|path|root).*'
\end{lstlisting}

To make the rule encoding concrete, we present two representative \tech rules,
one in each backend, covering one taxonomy category each. Both rules are
agent-agnostic in the sense that they target patterns shared by any local
LLM agent runtime that interpolates host paths into prompts or that joins
user-controlled segments into filesystem paths; neither rule depends on
OpenClaw-specific APIs or file paths.

\paragraph{Semgrep example: CAT-1 prompt injection via workspace path.}
Listing~\ref{lst:sg-prompt} encodes the pattern behind GHSA-2qj5-gwg2-xwc4
and a class of related advisories: a workspace, working directory, or project
root path is interpolated into a system prompt or prompt template without
sanitization. Because directory names can contain newline, Unicode bidi, or
zero-width characters, the prompt's instruction boundary can be broken from
outside the agent process. The rule combines a syntactic template (template
literal interpolating a variable) with a metavariable regular expression that
constrains the interpolated identifier's name to one of the canonical
path-bearing identifiers used across agent codebases. A
\texttt{pattern-not-inside} clause excludes calls that route through any
sanitizer.

\paragraph{CodeQL example: CAT-3 path traversal absence-of-guard.}
Listing~\ref{lst:cq-path} encodes a rule  
%Listing~\ref{lst:cq-path} shows a simplified version of the query.
that fires when a call to
\texttt{path.join(...)} receives a risky-named argument and no containment
check appears anywhere in the enclosing function. This rule illustrates the
expressiveness advantage discussed in Section~\ref{sec:rq2}: the
absence-of-guard predicate, ``no \texttt{startsWith}, \texttt{realpath}, or
similar check appears in the function containing the
\texttt{path.join}'', is expressed as a single existential over the function's
call set, including calls to helper functions that would be invisible to a
purely AST-local check---although it does not perform full interprocedural authorization reasoning. 
%The CodeQL rule encodes the missing-guard logic as a predicate over calls appearing in the enclosing function, which is more flexible than local AST templates but still bounded; it does not perform full interprocedural authorization reasoning. 
The rule's identifier-name regex is intentionally
broad so that it applies to any TypeScript or JavaScript runtime, not just
OpenClaw.

%\newpage
\begin{lstlisting}[style=qlstyle,caption={CAT-3 CodeQL query:
\texttt{path.join} called with a risky-named argument, with no containment
guard in the enclosing function.},label={lst:cq-path}]
/**
 * @name path.join with risky argument lacks containment check
 * @description path.join() is called with an argument whose identifier
 *   name suggests user-controlled input, and no startsWith() / realpath()
 *   / containment helper appears in the enclosing function.
 * @kind problem
 * @id agent/path-join-no-containment
 * @problem.severity warning
 * @tags security external/cwe/cwe-022
 */
import javascript

predicate hasContainmentGuard(Function f) {
  exists(InvokeExpr e |
    e.getEnclosingFunction() = f and
    e.getCalleeName() in [
      "startsWith", "isInside", "isWithin", "ensureInside",
      "realpath", "realpathSync", "resolveSafe", "assertWithinRoot"
    ]
  )
}

from MethodCallExpr c, Identifier arg
where
  c.getMethodName() = "join" and
  c.getReceiver().toString() = "path" and
  arg = c.getAnArgument() and
  arg.getName().regexpMatch(
    "(?i).*(input|param|body|query|req|user|external|untrusted|provided|client|filename|filepath|userpath|requested|target).*") and
  not hasContainmentGuard(c.getEnclosingFunction())
select c,
  "path.join() with risky argument '" + arg.getName() +
  "' has no containment check in its enclosing function."
\end{lstlisting}

\paragraph{Discussion.}
% The two listings expose the structural difference between the backends.
% The Semgrep rule encodes its missing-guard logic by enumerating known safe
% wrappers as \texttt{pattern-not-inside} clauses and by tightening the AST
% match around them; 
% %this scales linearly in the number of sanitizer names
% %and remains accurate when the sanitizer is close to the call site. 
% this works best when the sanitizer is close to the 
% matched call site. 
% The 
% CodeQL rule encodes the same missing-guard logic as a single existential
% over the function's call set, 
% %which scales to sanitizers that live in helper functions or library modules. 
% which scales to sanitizer calls made through helper functions, provided the guard call is visible in the enclosing function.
% %
% The two rules detect the same class
% of vulnerability but at different points on the expressiveness--scalability
% frontier, which is the basis of the cross-backend comparison reported in
% %Section~\ref{sec:rq4}. 
% Sections~\ref{sec:rq2} and~\ref{sec:rq5}.

The two listings expose the structural difference between the backends.
The Semgrep rule encodes missing-guard logic by enumerating known safe
wrappers as \texttt{pattern-not-inside} clauses and by tightening the AST
match around them; this works best when the sanitizer is close to the
matched call site. The CodeQL rule expresses the same idea as a predicate
over calls visible in the enclosing function, which makes the absence-of-guard
condition easier to state and combine with relational program facts. This
does not amount to full semantic authorization reasoning, but it illustrates
why the CodeQL backend can encode some guard patterns more compactly than
the Semgrep backend. This distinction helps explain the cross-backend
recall differences discussed in Section~\ref{sec:rq2}.

\section{\textsc{OpenClawBench} Construction Details}
\label{app:benchmark}

This appendix gives the implementation details behind
Section~\ref{sec:data-collection}.

\paragraph{Advisory fetching.}
We enumerated all advisories under the OpenClaw repository using the GitHub
Security Advisories REST endpoint with cursor-based pagination
(\texttt{/repos/openclaw/openclaw/security-advisories?per\_page=100} followed
by the \texttt{Link: rel="next"} cursor). For each of the 602 advisories
returned, we retrieved the JSON metadata exposed by the endpoint, including
the GHSA identifier, the CVE identifier (when assigned), the textual summary
and description, the severity label, the CVSS vector and score, the assigned
CWE class(es), the publication and update timestamps, the advisory state
(published, draft, withdrawn), and the structured affected-package list with
its vulnerable and patched version ranges. Advisories in the
\texttt{withdrawn} state were excluded.

\paragraph{Fix-commit recovery.}
The public Security Advisories endpoint does not return the SHA of the commit
that patched each advisory. We recovered it through OSV.dev, which mirrors the
GitHub Advisory Database and exposes \texttt{references} entries pointing at
\texttt{github.com/openclaw/openclaw/commit/<sha>} URLs. For each GHSA we
called the OSV API (\texttt{/v1/vulns/<ghsa-id>}) and extracted the
commit SHAs from its references. Of the 602 advisories, 365 had at least one
\texttt{commit/<sha>} reference. Among the remainder, some were either too recent for
OSV.dev's nightly sync (73), or carried no commit-typed reference in OSV (52),
or had a malformed commit reference (4). When multiple commits were listed,
we used the first one and verified that its parent existed; this matched the
manual fix-commit assignment in all spot checks.

\paragraph{Snapshot construction.}
For each advisory with a recovered fix commit, we identified the
\emph{vulnerable revision} as the first parent of the fix commit, then walked
the fix commit's file list (the \texttt{files[]} array of
\texttt{/repos/openclaw/openclaw/commits/<sha>}) and downloaded the
\emph{pre-fix} version of each modified file by fetching
\texttt{raw.githubusercontent.com/openclaw/openclaw/<parent-sha>/<path>}.
We preserved the repository directory structure and saved the result under
\texttt{vulnerable\_code/<ghsa-id>/}. We did not retrieve unrelated files
from the same revision; only those touched by the fix commit are included
in the snapshot, which keeps the average snapshot under 30~kB and avoids
contaminating per-advisory recall by inflating the file base.

\paragraph{File categorization.}
We labeled each file in a snapshot using its path: paths containing
\texttt{/test/}, \texttt{/tests/}, \texttt{/\_\_tests\_\_/},
\texttt{.test-d.ts}, or matching \texttt{*.test.*}/\texttt{*.spec.*} are
marked as test files; paths ending in \texttt{.md}, \texttt{.mdx},
\texttt{.rst}, \texttt{.adoc}, or \texttt{.txt} are marked as documentation;
all other files are stored as the vulnerable source code that the analysis
runs against. The categorization controls the denominator used in
Section~\ref{sec:rq1}: we report advisory-level recall only
over snapshots that contain at least one non-test, non-documentation file
in a scannable JavaScript or TypeScript extension. This filter excludes 26
advisories whose fix touched only Markdown, lockfiles, or JSON
configuration, yielding the 446 scannable advisories evaluated in the main
paper.

\paragraph{Temporal train/test split.}
We split the 446 scannable advisories by publication date with a cutoff of
\texttt{2026-04-01}. Of the 446, 229 were published before the cutoff and
form the \emph{rule-derivation set} (train); 217 were published on or after
the cutoff and form the \emph{held-out test set}. 
We chose this cutoff operationally: 
%We chose this cutoff because it was the date on which the rule-derivation corpus was frozen
it is the date on which we froze the working set of advisories
we had read while writing rules. Every advisory in the test set was therefore
collected after the rules were committed, eliminating any possibility of
inspecting an advisory's vulnerable file before encoding the rule that
catches it. We evaluated the test set exactly once, immediately before
preparing this manuscript.

\paragraph{HEAD live target.}
For the HEAD finding-volume measurements reported in
Section~\ref{sec:rq4}, we cloned the OpenClaw repository at commit
\texttt{4752e9a6} (2026-06-05), which contains 16{,}433 JavaScript or
TypeScript files (19{,}755 total files, 283~MB unpacked). We ran each tool
configuration against this single revision with default extractor settings.

\paragraph{Reproducibility.}
We publicly release \textsc{OpenClawBench} as a manifest of
\texttt{(ghsa\_id, fix\_sha, parent\_sha, file\_list, severity, cwe\_ids,
publication\_date, split)} tuples,  along with all 47 Semgrep YAML rules and
30 CodeQL queries. The benchmark can be rebuilt from the manifest using the
public GitHub and OSV APIs without further configuration.

\section{Additional Per-CWE Results}
\label{app:cwe}

This appendix extends Table~\ref{tab:cwe} from the main paper in two ways.
Table~\ref{tab:cwe-longtail} reports per-CWE recall for the next 14 CWE
classes (those with $n \in [5, 15]$ in \textsc{OpenClawBench}), and
Table~\ref{tab:cwe-traintest} reports the train/test split separately for
the top eight CWE classes so that the generalization claim in
Section~\ref{sec:rq3} can be inspected at finer granularity. 
%The results reveal three observations. 
The results support three observations.

\begin{table}[t]
\centering
\caption{Long-tail per-CWE recall on \textsc{OpenClawBench} ($n{=}446$, train+test union). CWE classes with $5 \le n \le 15$. The remaining $80+$ classes with $n \le 4$ are omitted; their combined size is comparable to CWE-863 alone but each contributes too few advisories for a stable per-CWE recall estimate. Inline annotations show the absolute change in pp vs.\ the corresponding baseline ($\uparrow$ increase, $\downarrow$ decrease).}
\label{tab:cwe-longtail}
\begin{tabular}{llccccc}
\toprule
 & & & \multicolumn{2}{c}{Semgrep (\%)} & \multicolumn{2}{c}{CodeQL (\%)} \\
\cmidrule(lr){4-5} \cmidrule(lr){6-7}
CWE & Description & $n$ & Pro & + \tech & Sec-Ext & + \tech \\
\midrule
CWE-59  & Symlink Following                   & 15 &  0.0 &  86.7\,\textcolor{green!55!black}{\scriptsize$\uparrow$86.7}  & 20.0 &  86.7\,\textcolor{green!55!black}{\scriptsize$\uparrow$66.7} \\
\rowcolor{gray!7}
CWE-200 & Information Exposure                & 10 &  0.0 &  40.0\,\textcolor{green!55!black}{\scriptsize$\uparrow$40.0}  & 10.0 &  80.0\,\textcolor{green!55!black}{\scriptsize$\uparrow$70.0} \\
CWE-639 & User-Controlled Resource Identifier & 10 & 40.0 &  50.0\,\textcolor{green!55!black}{\scriptsize$\uparrow$10.0}  &  0.0 &  60.0\,\textcolor{green!55!black}{\scriptsize$\uparrow$60.0} \\
\rowcolor{gray!7}
CWE-862 & Missing Authorization               & 10 & 10.0 &  50.0\,\textcolor{green!55!black}{\scriptsize$\uparrow$40.0}  &  0.0 &  40.0\,\textcolor{green!55!black}{\scriptsize$\uparrow$40.0} \\
CWE-15  & Environment Override                & 10 & 30.0 &  80.0\,\textcolor{green!55!black}{\scriptsize$\uparrow$50.0}  & 20.0 &  70.0\,\textcolor{green!55!black}{\scriptsize$\uparrow$50.0} \\
\rowcolor{gray!7}
CWE-269 & Improper Privilege Mgmt             &  9 & 11.1 &  44.4\,\textcolor{green!55!black}{\scriptsize$\uparrow$33.3}  &  0.0 &  66.7\,\textcolor{green!55!black}{\scriptsize$\uparrow$66.7} \\
CWE-693 & Protection Mechanism Failure        &  8 & 25.0 &  75.0\,\textcolor{green!55!black}{\scriptsize$\uparrow$50.0}  & 50.0 & 100.0\,\textcolor{green!55!black}{\scriptsize$\uparrow$50.0} \\
\rowcolor{gray!7}
CWE-426 & Untrusted Search Path               &  6 & 16.7 &  66.7\,\textcolor{green!55!black}{\scriptsize$\uparrow$50.0}  & 16.7 &  33.3\,\textcolor{green!55!black}{\scriptsize$\uparrow$16.6} \\
CWE-307 & Improper Throttling                 &  6 &  0.0 &  66.7\,\textcolor{green!55!black}{\scriptsize$\uparrow$66.7}  &  0.0 &  83.3\,\textcolor{green!55!black}{\scriptsize$\uparrow$83.3} \\
\rowcolor{gray!7}
CWE-266 & Incorrect Privilege Assign.         &  6 &  0.0 &  50.0\,\textcolor{green!55!black}{\scriptsize$\uparrow$50.0}  & 33.3 &  66.7\,\textcolor{green!55!black}{\scriptsize$\uparrow$33.4} \\
CWE-829 & Untrusted Functional Incl.          &  6 & 16.7 &  66.7\,\textcolor{green!55!black}{\scriptsize$\uparrow$50.0}  &  0.0 &  33.3\,\textcolor{green!55!black}{\scriptsize$\uparrow$33.3} \\
\rowcolor{gray!7}
CWE-770 & Resource Exhaustion                 &  6 & 66.7 &  83.3\,\textcolor{green!55!black}{\scriptsize$\uparrow$16.6}  & 16.7 & 100.0\,\textcolor{green!55!black}{\scriptsize$\uparrow$83.3} \\
CWE-290 & Spoofing Authentication             &  6 & 66.7 & 100.0\,\textcolor{green!55!black}{\scriptsize$\uparrow$33.3}  & 16.7 & 100.0\,\textcolor{green!55!black}{\scriptsize$\uparrow$83.3} \\
\rowcolor{gray!7}
CWE-436 & Interpretation Conflict             &  5 & 20.0 &  80.0\,\textcolor{green!55!black}{\scriptsize$\uparrow$60.0}  & 40.0 & 100.0\,\textcolor{green!55!black}{\scriptsize$\uparrow$60.0} \\
CWE-287 & Improper Authentication             &  5 & 60.0 &  80.0\,\textcolor{green!55!black}{\scriptsize$\uparrow$20.0}  & 40.0 &  80.0\,\textcolor{green!55!black}{\scriptsize$\uparrow$40.0} \\
\bottomrule
\end{tabular}
\end{table}

% \begin{table}[t]
% \centering
% \caption{Per-CWE recall split separately by train (rule-derivation, before
% 2026-04-01) and test (held-out, on or after 2026-04-01). Reported for the
% top eight CWE classes by frequency. For each backend, columns show
% $\text{train}\%$ / $\text{test}\%$. 
% %Small or negative gaps confirm that the custom rules do not overfit to the train advisories.
% Small or negative gaps are consistent with the custom rules capturing recurring structures rather than memorizing the rule-derivation advisories. 
% }
% \label{tab:cwe-traintest}
% \begin{tabular}{lcccccc}
% \toprule
%  & & \multicolumn{2}{c}{Semgrep tr/te (\%)} & \multicolumn{2}{c}{CodeQL tr/te (\%)} \\
% \cmidrule(lr){3-4} \cmidrule(lr){5-6}
% CWE & $n_{\text{tr}}/n_{\text{te}}$ & Pro & + \tech & Sec-Ext & + \tech \\
% \midrule
% CWE-863 & 40/37 & 30/16 &  48/70 &  5/14 & 80/81 \\
% CWE-22  & 24/7  &  8/43 &  79/100 & 17/14 & 79/86 \\
% CWE-78  & 20/10 & 50/30 &  80/60 & 25/10 & 90/60 \\
% CWE-284 & 11/10 & 27/30 &  64/100 &  9/10 & 91/80 \\
% CWE-918 & 13/7  & 23/29 &  77/71 & 15/0  & 77/86 \\
% CWE-285 & 11/7  & 27/0  &  64/71 &  0/0  & 82/86 \\
% CWE-400 & 14/3  & 36/100 & 64/100 & 14/33 & 86/100 \\
% CWE-184 &  9/7  & 22/0  &  22/57 & 11/14 & 44/57 \\
% \bottomrule
% \end{tabular}
% \end{table}

\begin{table}[t]
\centering
\caption{Per-CWE recall split separately by train (rule-derivation, before 2026-04-01) and test (held-out, on or after 2026-04-01). Reported for the top eight CWE classes by frequency. For each backend, columns show $\text{train}\%$ / $\text{test}\%$. Inline annotations on the \tech columns show the test$-$train change in pp ($\uparrow$: test exceeds train; $\downarrow$: test below train). Small magnitudes in either direction are consistent with custom rules capturing recurring structures rather than memorizing rule-derivation advisories.}
\label{tab:cwe-traintest}
\begin{tabular}{lccccc}
\toprule
 & & \multicolumn{2}{c}{Semgrep tr/te (\%)} & \multicolumn{2}{c}{CodeQL tr/te (\%)} \\
\cmidrule(lr){3-4} \cmidrule(lr){5-6}
CWE & $n_{\text{tr}}/n_{\text{te}}$ & Pro & + \tech & Sec-Ext & + \tech \\
\midrule
CWE-863 & 40/37 & 30/16  &  48/70\,\textcolor{green!55!black}{\scriptsize$\uparrow$22}  &  5/14 &  80/81\,\textcolor{green!55!black}{\scriptsize$\uparrow$1}  \\
\rowcolor{gray!7}
CWE-22  & 24/7  &  8/43  &  79/100\,\textcolor{green!55!black}{\scriptsize$\uparrow$21} & 17/14 &  79/86\,\textcolor{green!55!black}{\scriptsize$\uparrow$7}  \\
CWE-78  & 20/10 & 50/30  &  80/60\,\textcolor{red!55!black}{\scriptsize$\downarrow$20}    & 25/10 &  90/60\,\textcolor{red!55!black}{\scriptsize$\downarrow$30}  \\
\rowcolor{gray!7}
CWE-284 & 11/10 & 27/30  &  64/100\,\textcolor{green!55!black}{\scriptsize$\uparrow$36} &  9/10 &  91/80\,\textcolor{red!55!black}{\scriptsize$\downarrow$11}  \\
CWE-918 & 13/7  & 23/29  &  77/71\,\textcolor{red!55!black}{\scriptsize$\downarrow$6}    & 15/0  &  77/86\,\textcolor{green!55!black}{\scriptsize$\uparrow$9}  \\
\rowcolor{gray!7}
CWE-285 & 11/7  & 27/0   &  64/71\,\textcolor{green!55!black}{\scriptsize$\uparrow$7}   &  0/0  &  82/86\,\textcolor{green!55!black}{\scriptsize$\uparrow$4}  \\
CWE-400 & 14/3  & 36/100 &  64/100\,\textcolor{green!55!black}{\scriptsize$\uparrow$36} & 14/33 &  86/100\,\textcolor{green!55!black}{\scriptsize$\uparrow$14} \\
\rowcolor{gray!7}
CWE-184 &  9/7  & 22/0   &  22/57\,\textcolor{green!55!black}{\scriptsize$\uparrow$35}  & 11/14 &  44/57\,\textcolor{green!55!black}{\scriptsize$\uparrow$13} \\
\bottomrule
\end{tabular}
\end{table}

First, the agent-runtime CWEs that
are absent or minimally represented in the conventional web-security
literature (CWE-59 symlink following, CWE-15 environment override, CWE-693
protection-mechanism failure, CWE-290 spoofing authentication) all reach
80\%+ recall under \tech in at least one backend, and several reach 100\%.

Second, the two CWE classes the main paper flagged as residually difficult
(CWE-184 incomplete denylist and CWE-200 information exposure) both move
materially under the CodeQL backend, from 12\%/10\% baseline to 50\%/80\%
with \tech. They remain harder than the average CWE but are no longer
``baseline'' classes for which the static toolchain is silent. 

Third, the
finer-grained train/test split in Table~\ref{tab:cwe-traintest} shows that
the train-to-test recall gap is mostly small in absolute value and changes
sign across CWE classes (\emph{e.g.}, CWE-863 improves on test, CWE-78
regresses on test). This pattern is consistent with rules capturing genuine
syntactic structure rather than memorizing the train set.

Although generic Denial-of-Service behavior is outside our taxonomy scope, CWE-400 and CWE-770 advisories are retained when the advisory fix exposes a statically visible implementation weakness, such as an unbounded parser, missing size check, or unchecked allocation.

\section{STRIDE Mapping and Trust-Boundary Details}
\label{app:threat-model}

This appendix expands Section~\ref{sec:taxonomy-stride} with the full
STRIDE-to-category derivation. Table~\ref{tab:stride-full} lists, for each
STRIDE threat we kept, the runtime data-flow boundary at which it manifests,
the attacker-controlled data that crosses that boundary, the privileged
operation downstream, the runtime component that mediates the transition,
and the resulting \tech category. We also list the STRIDE threats we
\emph{excluded} together with the reason for exclusion.

\begin{table}[t]
\centering
\caption{Full STRIDE-to-category derivation. Each row lists a STRIDE threat,
the runtime boundary $B_i$ at which it manifests
(Section~\ref{sec:taxonomy-boundaries}), the data crossing the boundary,
the downstream privileged operation, the mediating runtime component, and
the resulting \tech category. The bottom of the table lists STRIDE threats
we excluded.}
\label{tab:stride-full}
\small
\begin{tabular}{p{1.8cm}p{0.3cm}p{3.0cm}p{4.0cm}p{3.6cm}p{1.2cm}}
\toprule
STRIDE threat & $B_i$ & Cross-boundary data & Downstream operation &
Mediating component & Category \\
\midrule
Tampering &
$B_1$ &
untrusted text destined for the prompt &
the model treats text as trusted instruction &
prompt builder, memory writer &
CAT-1 \\
\rowcolor{gray!7}
Tampering &
$B_2$ &
model- or user-controlled operand &
exec / spawn / eval / dynamic require &
tool dispatcher, skill loader &
CAT-2 \\
Elevation of Privilege &
$B_3$ &
filesystem path or sandbox arg &
resource access on host &
fs adapter, sandbox manager &
CAT-3 \\
\rowcolor{gray!7}
Information Disclosure &
$B_4$ &
URL, log message, or error &
outbound HTTP/WS, log emission, error formatter &
network client, logger &
CAT-4 \\
Spoofing &
$B_5$ &
caller identity at handler entry &
handler-side privileged action &
permission gate, route mount &
CAT-5 \\
\midrule
\multicolumn{6}{l}{\textit{Excluded threats:}} \\
\hline
Denial of Service & \multicolumn{5}{l}{\shortstack{Generic runtime-only DoS is out of scope; statically visible resource-exhaustion patterns, such as\\missing size checks or unbounded parsing, are retained under the closest implementation boundary.}} \\
\hline
Repudiation & \multicolumn{5}{l}{Concerns audit-logging integrity; orthogonal to the implementation weaknesses we target.} \\
\bottomrule
\end{tabular}
\end{table}

\paragraph{Why Tampering splits into two categories.}
We split Tampering into CAT-1 (prompt-side) and CAT-2 (execution-side)
because the attacker-controlled data crosses two structurally different
boundaries with two different mediating components: $B_1$ at the prompt
builder and $B_2$ at the tool dispatcher / skill loader. A rule that
guards $B_1$ (e.g., escapes control characters before placement in a system
prompt) does not guard $B_2$ (where the operand has already been emitted
by the model), and vice versa. Treating them as one category would either
hide the structural difference or force the auditor to write
double-rooted rules that do not factor cleanly. The taxonomy in
Section~\ref{sec:taxonomy} therefore keeps the two boundaries as separate
categories. The MECE (mutually exclusive and collectively exhaustive) check then reduces to verifying that every rule's
fix lives at exactly one $B_i$.

% \paragraph{Runtime data-flow diagram.}
% Figure~\ref{fig:dfd} (omitted; see supplementary materials) depicts the
% five boundaries $B_1$--$B_5$ within the OpenClaw agent runtime. Untrusted
% input enters at the top (user message, retrieved document, prior memory,
% tool output), traverses $B_1$ into the prompt builder, is consumed by the
% model, and exits the model as a structured action. The action traverses
% $B_2$ into the tool dispatcher and onwards through $B_3$ to the host
% filesystem or sandbox, or through $B_4$ to the network. Caller identity
% enters at the handler edge through $B_5$. Each \tech rule names the
% boundary it guards in its \texttt{metadata.cat} field, which we used to
% mechanically check the MECE property during rule development.

\paragraph{Runtime data-flow diagram.} Untrusted
input enters at the top (user message, retrieved document, prior memory,
tool output), traverses $B_1$ into the prompt builder, is consumed by the
model, and exits the model as a structured action. The action traverses
$B_2$ into the tool dispatcher and onwards through $B_3$ to the host
filesystem or sandbox, or through $B_4$ to the network. Caller identity
enters at the handler edge through $B_5$. Each \tech rule names the
boundary it guards in its \texttt{metadata.cat} field, which we used to
mechanically check the MECE property during rule development.

\end{document}